\documentclass[twocolumn,superscriptaddress,nofootinbib,amssymb,aps,prd,floatfix]{revtex4-2}

\usepackage{orcidlink}
\usepackage{mathtools}
\usepackage{hyperref}
\usepackage{cleveref}
\crefname{equation}{Eq.\@}{Eq.s\@}
\Crefname{equation}{Equation}{Equations}
\crefname{table}{Table}{Tables}
\Crefname{table}{Table}{Tables}
\crefname{figure}{Fig.\@}{Fig.s\@}
\Crefname{figure}{Figure}{Figures}
\crefname{section}{Section}{Sections}
\Crefname{section}{Section}{Sections}
\AddToHook{cmd/appendix/before}{\crefalias{section}{appendix}}
\crefname{appendix}{Appendix}{Appendices}

\usepackage{siunitx}
\usepackage{graphicx}
\usepackage{longtable}
\usepackage{siunitx}
\usepackage[normalem]{ulem}
\usepackage{booktabs}

\newcommand{\eg}{e.\,g.\@ }
\newcommand{\ie}{i.\,e.\@}
\newcommand{\pardev}[2]{\frac{\partial #1}{\partial #2}}

\newcommand*{\defeq}{\mathrel{\vcenter{\baselineskip0.5ex \lineskiplimit0pt
                     \hbox{\scriptsize.}\hbox{\scriptsize.}}}%
                     =}
\newcommand*{\boldone}{\text{\usefont{U}{bbold}{m}{n}1}}
\newcommand*\diff{\mathop{}\!\mathrm{d}}

\DeclareMathOperator*{\argmin}{arg\,min}   
\DeclareMathOperator*{\argmax}{arg\,max}

\newcommand{\signal}{\ensuremath{\mathrm{S}}}
\newcommand{\candidate}{\ensuremath{\mathrm{c}}}

\newcommand{\data}{\ensuremath{d}}

\newcommand{\params}{\ensuremath{\omega}}
\newcommand{\dparams}{\ensuremath{\vartheta}}
\newcommand{\aparams}{\ensuremath{\mathcal{A}}}

\newcommand{\likelihood}{\ensuremath{\mathcal{L}}}

\newcommand{\searchspace}{\ensuremath{\mathcal{R}}}

\newcommand{\fstat}{\ensuremath{\mathcal{F}}}

\newcommand{\measurement}{D}
\newcommand{\dataincoherent}{\ensuremath{\{\data_\ell\}}}

\newcommand{\hwinjreftimestageseven}{1253764756}

\newcommand{\hwinjzerofreq}{\num{265.57502155}}
\newcommand{\hwinjzerofonedot}{\num{-4.15e-12}}
\newcommand{\hwinjzeroalpha}{\ensuremath{4{:}46{:}12.4628}}
\newcommand{\hwinjzerodelta}{\ensuremath{-56{:}13{:}02.9490}}
\newcommand{\hwinjzerodfreq}{\num{6.5e-08}}
\newcommand{\hwinjzerodfonedot}{\num{9.0e-15}}
\newcommand{\hwinjzerodsky}{\ensuremath{0{:}00{:}00.4012}}

\newcommand{\hwinjtwofreq}{\num{575.16350421}}
\newcommand{\hwinjtwofonedot}{\num{-1.37e-13}}
\newcommand{\hwinjtwoalpha}{\ensuremath{14{:}21{:}01.4800}}
\newcommand{\hwinjtwodelta}{\ensuremath{3{:}26{:}38.3626}}
\newcommand{\hwinjtwodfreq}{\num{3.6e-08}}
\newcommand{\hwinjtwodfonedot}{\num{4.3e-15}}
\newcommand{\hwinjtwodsky}{\ensuremath{0{:}00{:}00.2198}}

\newcommand{\hwinjthreefreq}{\num{108.8571594}}
\newcommand{\hwinjthreefonedot}{\num{-1.46e-17}}
\newcommand{\hwinjthreealpha}{\ensuremath{11{:}53{:}29.4178}}
\newcommand{\hwinjthreedelta}{\ensuremath{-33{:}26{:}11.7687}}
\newcommand{\hwinjthreedfreq}{\num{1.8e-07}}
\newcommand{\hwinjthreedfonedot}{\num{2.4e-14}}
\newcommand{\hwinjthreedsky}{\ensuremath{0{:}00{:}02.9579}}

\newcommand{\hwinjfivefreq}{\num{52.80832436}}
\newcommand{\hwinjfivefonedot}{\num{-4.03e-18}}
\newcommand{\hwinjfivealpha}{\ensuremath{20{:}10{:}30.3939}}
\newcommand{\hwinjfivedelta}{\ensuremath{-83{:}50{:}20.9036}}
\newcommand{\hwinjfivedfreq}{\num{4.8e-08}}
\newcommand{\hwinjfivedfonedot}{\num{6.0e-15}}
\newcommand{\hwinjfivedsky}{\ensuremath{0{:}00{:}01.6488}}

\newcommand{\hwinjninefreq}{\num{763.84731649}}
\newcommand{\hwinjninefonedot}{\num{-1.45e-17}}
\newcommand{\hwinjninealpha}{\ensuremath{13{:}15{:}32.5397}}
\newcommand{\hwinjninedelta}{\ensuremath{75{:}41{:}22.5205}}
\newcommand{\hwinjninedfreq}{\num{-7.9e-08}}
\newcommand{\hwinjninedfonedot}{\num{-9.5e-15}}
\newcommand{\hwinjninedsky}{\ensuremath{0{:}00{:}00.1649}}

\newcommand{\hwinjtenfreq}{\num{26.33144237}}
\newcommand{\hwinjtenfonedot}{\num{-8.5e-11}}
\newcommand{\hwinjtenalpha}{\ensuremath{14{:}46{:}13.3549}}
\newcommand{\hwinjtendelta}{\ensuremath{42{:}52{:}38.2953}}
\newcommand{\hwinjtendfreq}{\num{5.7e-08}}
\newcommand{\hwinjtendfonedot}{\num{8.3e-15}}
\newcommand{\hwinjtendsky}{\ensuremath{0{:}00{:}03.9060}}

\NewDocumentCommand{\lratio}{ e{^} s o >{\SplitArgument{1}{|}}m }{%
    \operatorname{\Lambda}
    \IfValueT{#1}{{\!}^{#1}}
    \IfBooleanTF{#2}{
        \expectarg*{\expectvar#4}%
    }{
        \IfNoValueTF{#3}{
            \expectarg{\expectvar#4}%
        }{
            \expectarg[#3]{\expectvar#4}%
        }%
    }%
}
\NewDocumentCommand{\lik}{ e{^} s o >{\SplitArgument{1}{|}}m }{%
    \operatorname{\mathcal{L}}
    \IfValueT{#1}{{\!}^{#1}}
    \IfBooleanTF{#2}{
        \expectarg*{\expectvar#4}%
    }{
        \IfNoValueTF{#3}{
            \expectarg{\expectvar#4}%
        }{
            \expectarg[#3]{\expectvar#4}%
        }%
    }%
}

\NewDocumentCommand{\probdensity}{ e{_} e{^} s o >{\SplitArgument{1}{|}}m }{%
    \IfValueTF{#1}{f_{#1}}{p}
    \IfValueT{#2}{^{#2}}
    \IfBooleanTF{#3}{
        \expectarg*{\expectvar#5}%
    }{
        \IfNoValueTF{#4}{
            \expectarg{\expectvar#5}%
        }{
            \expectarg[#4]{\expectvar#5}%
        }%
    }%
}

\NewDocumentCommand{\prob}{ e{^} s o >{\SplitArgument{1}{|}}m }{%
    \operatorname{P}
    \IfValueT{#1}{{\!}^{#1}}
    \IfBooleanTF{#2}{
        \expectarg*{\expectvar#4}%
    }{
        \IfNoValueTF{#3}{
            \expectarg{\expectvar#4}%
        }{
            \expectarg[#3]{\expectvar#4}%
        }%
    }%
}

\NewDocumentCommand{\gaussian}{ e{^} s o >{\SplitArgument{1}{|}}m }{%
    \mathcal{G}
    \IfValueT{#1}{{\!}^{#1}}
    \IfBooleanTF{#2}{
        \expectarg*{\expectvar#4}%
    }{
        \IfNoValueTF{#3}{
            \expectarg{\expectvar#4}%
        }{
            \expectarg[#3]{\expectvar#4}%
        }%
    }%
}

\NewDocumentCommand{\unif}{ e{^} s o >{\SplitArgument{1}{|}}m }{%
    \mathcal{U}
    \IfValueT{#1}{{\!}^{#1}}
    \IfBooleanTF{#2}{
        \expectarg*{\expectvar#4}%
    }{
        \IfNoValueTF{#3}{
            \expectarg{\expectvar#4}%
        }{
            \expectarg[#3]{\expectvar#4}%
        }%
    }%
}

\NewDocumentCommand{\cdf}{ e{^} s o >{\SplitArgument{1}{|}}m }{%
    \operatorname{CDF}
    \IfValueT{#1}{{\!}^{#1}}
    \IfBooleanTF{#2}{
        \expectarg*{\expectvar#4}%
    }{
        \IfNoValueTF{#3}{
            \expectarg{\expectvar#4}%
        }{
            \expectarg[#3]{\expectvar#4}%
        }%
    }%
}
\NewDocumentCommand{\fstatistic}{ e{_} e{^} s o >{\SplitArgument{1}{|}}m }{%
    \operatorname{\mathcal{F}}
    \IfValueT{#1}{{\!}_{#1}}
    \IfValueT{#2}{{\!}^{#2}}
    \IfBooleanTF{#3}{
        \expectarg*{\expectvar#5}%
    }{
        \IfNoValueTF{#4}{
            \expectarg{\expectvar#5}%
        }{
            \expectarg[#4]{\expectvar#5w}%
        }%
    }%
}
\NewDocumentCommand{\semcohfstatistic}{ e{^} s o >{\SplitArgument{1}{|}}m }{%
    \bar{\mathcal{F}}
    \IfValueT{#1}{{}^{#1}}
    \IfBooleanTF{#2}{
        \expectarg*{\expectvar#4}%
    }{
        \IfNoValueTF{#3}{
            \expectarg{\expectvar#4}%
        }{
            \expectarg[#3]{\expectvar#4}%
        }%
    }%
}
\NewDocumentCommand{\expectvar}{mm}{%
    #1\IfValueT{#2}{\nonscript\;\delimsize\vert\nonscript\;#2}%
}
\DeclarePairedDelimiterX{\expectarg}[1]{(}{)}{#1}

\makeatletter

\def\xxhref#1 {%
\ifx\^#1\expandafter\@gobble\else\expandafter\@firstofone\fi
{\mbox{\href{\tmp}{#1}} \xxhref}}
\makeatother

\begin{document}


\title{Bayesian Framework to Follow-up Continuous Gravitational Wave Candidates from Deep Surveys}
\author{J.~Martins\,\orcidlink{0009-0002-3912-189X}}
\email{jasper.martins@aei.mpg.de}
\author{M.~A.~Papa\,\orcidlink{0000-0002-1007-5298}}
\author{B.~Steltner\,\orcidlink{0000-0003-1833-5493}}
\author{R.~Prix\,\orcidlink{0000-0002-3789-6424}}
\affiliation{Max Planck Institute for Gravitational Physics (Albert Einstein Institute), D-30167 Hannover, Germany}
\affiliation{Leibniz Universit\"at Hannover, D-30167 Hannover, Germany}
\author{P.~B.~Covas\,\orcidlink{0000-0002-1845-9309}}
\affiliation{Max Planck Institute for Gravitational Physics (Albert Einstein Institute), D-30167 Hannover, Germany}
\affiliation{Leibniz Universit\"at Hannover, D-30167 Hannover, Germany}
\affiliation{Departament de Física, Universitat de les Illes Balears, IAC3 – IEEC, Carretera Valldemossa km 7.5, E-07122 Palma, Spain}

\date{\today}
\begin{abstract}
  Broad all-sky searches for continuous gravitational waves have high computational costs and require hierarchical pipelines. The sensitivity of these approaches is set by the initial search and by the number of candidates from that stage that can be followed up.
  The current follow-up schemes for the deepest surveys require careful tuning and set-up, have a significant human-labor cost and this impacts the number of follow-ups that can be afforded.
  Here we present and demonstrate
  a new follow-up framework based on Bayesian parameter estimation for the rapid, highly automated follow-up of candidates produced by the early stages of deep, wide-parameter space searches for continuous waves.
\end{abstract}


\maketitle

\section{Introduction}\label{sec:introduction}

Gravitational waves emitted during the merger of two compact objects are now routinely detected in data gathered by the LIGO-Virgo-KAGRA Collaboration~\cite{KAGRA:2021vkt,LIGOScientific:2024elc}.

Among the gravitational wave signals that remain elusive are continuous gravitational waves emitted by fast-rotating neutron stars with a sustained quadrupole moment~\cite{Pagliaro:2023bvi,Riles:2022wwz}, as well as other more exotic scenarios~\cite{Zhu:2020tht,Arvanitaki:2014wva}.

The long duration of continuous waves renders signals detectable that are at any instant of time below the detector noise, and makes it possible to resolve different waveforms with great accuracy.
For this reason, all-sky searches over a broad frequency range using coherent detection methods are computationally not viable.
This has long been known~\cite{Brady:1997ji,Brady:1998nj,Schutz:1999mb,Krishnan:2004sv,Cutler:2005hc,Dergachev:2011pd} and hierarchical semi-coherent searches have proven to be the most sensitive solution to the problem~\cite{Steltner:2023cfk,KAGRA:2022dwb,Dergachev:2025ead,McGloughlin:2025eso,McGloughlin:2025iyx}.

The initial search in the hierarchy examines the full parameter space with an incoherent combination of the results of coherent searches over  short data segments, spanning less than one hour to several days, depending on the search.
The shorter coherence times reduce the required number of template waveforms and, hence, the computing cost, at the cost of a reduced sensitivity.
Low detection thresholds can be chosen to ease this effect, but result in a high number of false alarms.
These are weeded out with several follow-up stages that employ increasingly higher coherence times, finer resolutions, and more stringent detection thresholds and vetoes. The goal is to reject noise while ensuring that detectable gravitational-wave signals survive.

Each follow-up stage of a hierarchical search requires (i) the identification of signal candidates, typically as local maxima of a detection statistic or as the result of some clustering procedure; (ii) the definition of thresholds and vetoes that reject candidates; (iii) the determination of per-candidate parameter space regions around each candidate that are expected to contain an underlying signal; (iv) the definition of a setup of the next search stage.

The first hierarchical search with a significant follow-up effort  was carried out in~\cite{Papa:2016cwb} on the results of an Einstein@Home search. The latest generation Einstein@Home searches are characterized by long coherence-times -- possibly the longest among broad searches -- and the results are investigated with the largest follow-up campaigns involving millions of candidates~\cite{Steltner:2023cfk,Ming:2024duga}. This is about an order of magnitude more candidates than examined with other approaches: for example~\cite{KAGRA:2022dwb} followed up between $10^3$ and $10^5$ candidates, depending on the pipeline. The long coherence time and the low follow-up thresholds pose unique challenges to the follow-up schemes.

All  the initial Einstein@Home search stages~\cite{Steltner:2023cfk,Ming:2024duga}, as well as all initial stages in~\cite{KAGRA:2022dwb}, and all but one of the~\cite{KAGRA:2022dwb} follow-ups are completely deterministic:
The signal-parameter uncertainty region around each candidate at stage $n$ is completely covered with a template bank and systematically investigated at stage $n+1$. This  approach is warranted by two factors: 1) It is easiest to simply redo on a smaller parameter space what one has already done on a larger one; 2) in the first follow-up stages the putative signal of the long coherence-time large-scale follow-ups is still weak enough and/or the uncertainty in signal parameters is large enough that stochastic methods are thought to not perform sufficiently well.

On the other hand stochastic search methods have the potential to be more efficient than deterministic template-bank based methods, as they concentrate compute cycles on the interesting regions of the parameter space. This becomes crucial when inspecting high dimensional spaces, like those pertaining signals from unknown neutron stars with a companion.
The more efficient allocation of compute cycles also allows for a more precise parameter estimation of putative signals~\cite{Covas:2024pam}.

Another advantage of stochastic search methods is that there already exist tools to integrate Monte Carlo sampling techniques in Bayesian parameter estimation frameworks, which would provide a natural way to estimate the posterior probability density function for any parameter after a search.
Posterior parameter distributions would solve a short-coming of the current methods that evaluate the uncertainty regions conservatively from the distributions of distances between recovered signal parameters and the actual parameters of fake signals from the target population. This approach results in search-region extents that are the same for every candidate and are overestimated for the majority of them. Conversely, the posterior can be calculated specifically for every candidate and inform the search-sampling in the next stage.

Non-deterministic follow-ups of continuous wave candidates have been demonstrated for short coherence times (less than a day)~\cite{Tenorio:2021njf,KAGRA:2022dwb,LIGOScientific:2020qhb,Covas:2024nzs} and for long coherence times (several days) for signals louder than those targeted by the most sensitive all-sky searches~\cite{Covas:2024pam}.
The regime of the weakest signals searched with long coherence times~\cite{Dergachev:2025ead,Steltner:2023cfk} remains unexplored. Here, we present a new Bayesian framework for hierarchical searches that establishes nondeterministic follow-up capabilities in that regime, in a highly automated manner. We demonstrate our framework on real data, following up the Einstein@Home search candidates of~\cite{Steltner:2023cfk}, and find  results consistent with the findings of the previous deterministic search.

\Crefrange{sec:introduction}{sec:framework} motivate the development of the new framework, present the signal model, discuss the background of Bayesian inference and provide the specific context of Einstein@Home searches.

In \cref{sec:testingtheframework}, we present the results of applying the framework to the Einstein@Home~\cite{Steltner:2023cfk}. We assess the stage of the follow-up at which our framework becomes applicable. We follow up all candidates that remained at that stage in the original search, compare the required computational and human effort, and compare the final results.

\Cref{sec:conclusion} concludes the paper with a discussion of our findings, and an outlook on future research.

\section{Signal Model and \fstat-statistic}\label{sec:signalModelAndFstat}

We consider a continuous wave signal that at the detector takes the form~\cite{Jaranowski:1998qm}:
\begin{equation}
  h(t)=F_+(t)h_+(t)+F_{\times}(t)h_{\times}(t).
  \label{eq:signal}
\end{equation}
$F_+(t)$ and $F_\times(t)$ are  the detector beam-pattern functions for the two gravitational wave polarizations ``+'' and ``$\times$''.
They depend on the sky position of the source (defined by the ecliptic longitude $\lambda$ and
ecliptic latitude $\beta$), and the orientation $\psi$ of the wave-frame with respect to the detector frame.
The waveforms for the two polarizations, $h_+(t)$ and $h_\times(t)$, are
\begin{eqnarray}
  h_+ (t)  =  A_+ \cos \phi(t) \nonumber \\
  h_\times (t)  =  A_\times \sin \phi(t),
  \label{eq:monochromatic}
\end{eqnarray}
with  $\phi(t)$ being the phase of the gravitational-wave signal at the time $t$ and $A_{+,\times}$ the polarizations' amplitudes:
\begin{eqnarray}
  A_+  & = & {\frac{1}{2}} h_0 (1+\cos^2\iota) \nonumber \\
  A_\times & = &  h_0  \cos\iota.
  \label{eq:amplitudes}
\end{eqnarray}
Here $h_0$ is the intrinsic gravitational wave amplitude and $\iota$ is the angle between the total angular momentum of the star and the line of sight.

The phase $\Phi(\tau)$ of the signal for an observer at rest with respect to the source (for an isolated source this is typically the solar system barycenter) is:
\begin{multline}
  \label{eq:phiSSB}
  \Phi(\tau) = \Phi_0 + 2\pi [ f (\tau - \tau_0)  + {\frac{1}{2}} \dot{f} (\tau-\tau_0)^2 + \cdots ],
\end{multline}
where $f$ is the signal frequency and $\tau$ is the arrival time of the gravitational wave front at the solar system barycenter and $\tau_0$ is a reference time. The detector frame phase $\phi(t)$ incorporates the effects due to the relative motion between the source and the detector, as well as relativistic effects. $\phi(t) = \Phi(\tau(t))$, with the transformation $\tau(t)$ depending on the position of the source.

We will consider an isolated source with a signal model referred to as the IT1-model~\cite{Dergachev:2020upb}, \ie, a signal emitted by an isolated neutron with only one spindown. This model is described by two sets of parameters $\params\defeq ({\dparams,{A}})$ that define the
\begin{equation}
  \begin{matrix*}[l]
    \textrm{phase-evolution:} &\dparams=(\lambda,\beta,f,\dot{f})\\
    \textrm{and amplitude:}   &A=(h_0,\cos\iota,\psi,\Phi_0)
  \end{matrix*}
\end{equation}
in the detector frame.

The optimal Neyman-Pearson detection statistic for a known signal is the likelihood ratio $\Lambda$:
\begin{equation}  \label{eq:likelihoodRatio}
  \Lambda(d,w) = \frac{\probdensity_{\measurement|\mathrm{IT1}}{\measurement=\data|\params}}{\probdensity_{\measurement|\mathrm{IT1}}{\measurement=\data|0}},
\end{equation}
where $\probdensity_{\measurement|\mathrm{IT1}}{\measurement=\data|\cdot}$ is the probability density of the data $d$ under a Gaussian noise hypothesis and the IT1 signal model.

By a judicious reparametrization of the signal amplitudes, the signal waveform can be written as a linear combination of four basis functions that depend on the phase-evolution parameters $\dparams$ and time, with coefficients $\aparams^\mu$ that depend only on the physical amplitude parameters $A$~\cite{Jaranowski:1998qm}:
\begin{equation}
  h(t;\params) = \sum_{\mu=1}^4 \mathcal{A}^\mu(A) h_\mu(t;\dparams).
  \label{eq:signalLinearCombination}
\end{equation}
In this form, an analytic maximization of the likelihood ratio can be carried out over the $\aparams^\mu$. The resulting detection statistic is the $\fstat$-statistic~\cite{Jaranowski:1998qm, Cutler:2005hc}:
\begin{equation}
  \label{eq:fstatMLE}
  2\!\fstatistic{\dparams} \defeq 2 \log\Lambda(\data,\aparams_\mathrm{MLE}, \dparams),
\end{equation}
where $\aparams_\textrm{MLE} = \argmax_\aparams \Lambda(\data, \aparams, \dparams)$, which implicitly depends on the phase-evolution parameters.

In Gaussian noise, the $2\fstat$-statistic is a sum of four Gaussian random variables and, as such, follows a $\chi^2$-distribution with four degrees of freedom and noncentrality parameter $\rho^2(\params_{\signal},\dparams_t)$ that depends on the signal parameters $\params_{\signal} $ and on the template parameters $\dparams_t$. 
In the perfectly matched case $(\dparams_t = \dparams_{\signal})$, the noncentrality parameter is~\cite{Dreissigacker:2018afk}
\begin{equation}
  \label{eq:rho2perfectmatch}
  \rho^2(\params_{\signal},\dparams_{\signal}) = \frac{4}{25}\frac{T_\mathrm{data}}{S_n} h_0^2 R(\lambda, \beta, \psi, \cos\iota),
\end{equation}
where $T_\mathrm{data}$ is the total data duration, $S_n$ is the single-sided noise power spectral density at the signal frequency, and $R(\cdot)$ describes the sky position and polarization dependent detector response.
Given a template with parameters $\dparams_t=\dparams_{\signal} + \Delta \dparams$, the noncentrality parameter is lower than for a perfectly matched template:
\begin{equation}
  \label{eq:rho2loss}
  \rho^2(\params_{\signal},\dparams_{\signal} + \Delta \dparams)\leq \rho^2(\params_{\signal},\dparams_{\signal}).
\end{equation}
When the distance between the signal and the template is small, the relative loss
is well approximated by a quadratic function of the distance:
\begin{equation}
  \label{eq:mismatch}
  \frac{\rho^2(\params_{\signal},\dparams_{\signal})-\rho^2(\params_{\signal},\dparams_{\signal} + \Delta \dparams)}{\rho^2(\params_{\signal},\dparams_{\signal})}\simeq g_{ij}(\params_{\signal}) \Delta \dparams^i \Delta \dparams^j,
\end{equation}
with
\begin{equation}
  \label{eq:metric}
  g_{ij}(\params_{\signal}) \defeq \left.\frac{-1}{2\rho^2(\params_{\signal},\dparams_{\signal})}\pardev{^2\rho^2(\params_{\signal},\dparams_\mathrm{t})}{\dparams^i_\mathrm{t}~\partial\dparams^j_\mathrm{t}}\right|_{\dparams_\mathrm{t}=\dparams_{\signal}}.
\end{equation}
\Cref{eq:metric} gives the so-called $\fstat$-statistic metric~\cite{Prix:2006wm}.
Typically, approximations that neglect the dependency on the amplitude parameters $\aparams_{\signal}$ are used~\cite{Prix:2006wm,Wette:2015lfa}.
We use such an approximation, the phase metric, defined as
\begin{align}\label{eq:phasemetric}
  \tilde g_{ij}(\dparams_{\signal}) \defeq & \left[ \left\langle  \pardev{\phi(t, \dparams)}{\dparams_i} \pardev{\phi(t,\dparams)}{\dparams_j} \right\rangle_T \right.\nonumber                                                   \\
                                           & \left.- \left\langle \pardev{\phi(t, \dparams)}{\dparams_i}\right\rangle_T \left\langle \pardev{\phi(t,\dparams)}{\dparams_j} \right\rangle_T \right]_{\dparams=\dparams_{\signal}},
\end{align}
where $\langle\cdot\rangle_{T}$ represents an average over the observing time.

Semi-coherent searches split the data into $N_\mathrm{seg}$ segments $\{\data_\ell\}$. The semi-coherent $\fstat$-statistic is the (averaged) joint log-likelihood ratio of signals with the same phase-evolution parameters $\dparams$ in all $N_\mathrm{seg}$ data segments, but independently maximized over the amplitude parameters for each segment~\cite{Cutler:2005hc}:
\begin{align}\label{eq:semcohfstatistic}
  2\semcohfstatistic{\dparams} & \defeq \frac{1}{N_\mathrm{seg}}\sum_{\ell=1}^{N_\mathrm{seg}}2\log\Lambda(\data_\ell, \aparams_{\mathrm{MLE},\ell}, \dparams).
\end{align}
The $\aparams_{\mathrm{MLE},\ell}$ are the maximum likelihood estimators for each segment.
We will only use segmentations where all segments have the same length, $T_\mathrm{coh}$.

In Gaussian noise, $N_\mathrm{seg}2\bar\fstat$ follows a $\chi^2$-distribution with $4N_\mathrm{seg}$ degrees of freedom and a noncentrality parameter that is the sum of all $\rho^2_\ell$.
The semi-coherent $\fstat$-statistic phase metric is the weighted average of the metrics for each data segment~\cite{covas2025}. We neglect the weighting operation. We use the semi-coherent metric in \cref{sec:posts_and_envs} when modeling the posterior probability density function of signal candidates.

\section{Bayesian inference scheme for continuous wave candidate follow-up}\label{sec:framework}

\subsection{Notation}

In this paper, we adopt the following notation for random variables. If not specified otherwise, all random variables discussed are continuous.
For any continuous random variable $X$ with probability density function $f_{X}$, we define
\begin{equation}\label{eq:notationdensity}
  \probdensity{x} \defeq \probdensity_{X}{X=x}.
\end{equation}
Similarly, if $X$ conditionally depends on the value taken by another continuous random variable $Y$ and some other information $I$, we define
\begin{equation}\label{eq:notationconditionaldensity}
  \probdensity{x|y, I} \defeq \probdensity_{X|Y, I}{X=x|Y=y, I}.
\end{equation}
We also refer to $\probdensity{x}$ just as the probability of $x$.

\subsection{Background}
Bayesian statistic allows to make inferences on the values of the parameters $\params$ affecting the outcome of a measurement $\data$ in the presence of noise.
The measuring device and the noise are described in terms of $\probdensity{\data|\params}$, the probability to measure $\data$ in the presence of a signal with parameter values $\params$ -- this is the probability of $\data$ \emph{conditioned} upon $\params$.
The inference is made by deriving the probability distribution function for $\params$ given the measurement $\data$:  $\probdensity{\params|\data}$, the so-called posterior distribution.
Bayes' theorem tells us how to invert the order of conditioning and derive the posterior $\probdensity{\params
    |\data}$, given the probability of $\probdensity{\data|\params}$ and our prior beliefs of the probability of the parameters, $\probdensity{\params}$.
Here, where both the parameters and the data take continuous values, Bayes' theorem takes the form~\cite{gelmanBayesianDataAnalysis2014}:
\begin{equation}
  \label{eq:parambayes}
  \probdensity{\params|\data} = \frac{\probdensity{\data|\params} \probdensity{\params}}{\int_\Omega \probdensity{\data|\params} \probdensity{\params} \diff \params},
\end{equation}
where $\Omega$ indicates the sample space of the parameters: $\params \in \Omega$.

The denominator in \cref{eq:parambayes},
\begin{equation}\label{eq:bayesevidence}
  \probdensity{\data|\Omega} = \int_{\Omega} \probdensity{\data |\params}\probdensity{\params}\diff \params,
\end{equation}
is called the \emph{evidence} and measures the support for the hypothesis of a signal with parameter values $\params \in \Omega$.
The ratio of the evidences for the signal to the null-hypothesis of Gaussian noise gives the Bayes' factor
\begin{align}\label{eq:bayesfactor}
  \mathcal{B}_{\signal/\mathrm{G}} = \frac{\probdensity{\data|\Omega}}{\probdensity{\data|0}}.
\end{align}
In Bayesian statistics different models (or hypotheses) are compared via Bayes' factors.
A measurement that yields a Bayes' factor with $\mathcal{B} > 1$ favors the signal hypothesis~\cite{gelmanBayesianDataAnalysis2014}.

\Cref{eq:parambayes,eq:bayesfactor} have historically
been used to search for continuous waves from pulsars with well-known phase-evolution parameters~\cite{LIGOScientific:2021hvc}.
In those searches the final product is the posterior distribution for $h_0$, marginalized over all other unknown signal parameters. A highly peaked posterior, far from zero, would be the first indication of the detection of a signal.

Broad surveys have typically taken a frequentist approach based on the $\fstat$-statistic. The main advantage of this method is that it maximizes the likelihood ratio over the amplitude parameters $\aparams$, hence does not require an explicit search over them and this improves the computational efficiency.  The parameters that remain for a search are the phase-evolution parameters $\dparams$.

In recent years, Bayesian frameworks have also utilized the $\fstat$-statistic likelihood~\cite{Ashton:2018ure}, for candidate follow-ups~\cite{Tenorio:2021njf,Covas:2024pam} and for parameter estimation~\cite{Ashok:2024fts}. 
A significant advantage of this approach is that the $\fstat$-statistic calculations have been highly optimized and are hence very efficient, with a number of tools publicly available~\cite{lalsuite,Wette:2020air}.
The work presented here is in this vein.

The first stage -- often referred to as Stage 0 -- of a large survey is the most challenging because the whole parameter space is searched. Only the parameter space regions associated with ``interesting" results are ear-marked for further inspection.
These are the so-called candidates that are followed up with the hierarchy of semi-coherent searches.
Typically, the candidates are identified by some clustering procedure and detection statistic threshold, and represented by a point in parameter space - the \emph{seed} - and a parameter uncertainty region around it (see e.g.~\cite{Steltner:2023cfk,Steltner:2022aze}).
On a large scale production run, no stochastic method has yet been demonstrated competitive with this sort of approach~\cite{Beheshtipour:2020zhb,Beheshtipour:2020nko}.

We shall hence assume that a follow-up using the new framework begins with a set of given candidates, for which we know what the parameter uncertainties are. We will also assume that the coherence time hierarchy of the various semi-coherent searches is given.

For every candidate we run a succession of follow-ups. At every follow-up search we derive the posteriors on the signal parameters and the evidence for the presence of a signal.
We use the former to define the search at the next stage. At the end of the follow-ups, we use the evidences to decide whether to discard the candidate or not~\cite{Searle:2008jv}.
The remainder of this section describes these steps in more detail.

\subsection{Using the \fstat-statistic for Bayesian inference}\label{sec:BayesianFstat}

Our starting point is a list of candidates and their equal-size phase-evolution parameters uncertainty. For each candidate the follow-up infers the phase-evolution parameters $\dparams$ of a putative signal in that range. For every candidate each follow-up stage produces i) the posterior distribution for the $\dparams$ that will be used to define the search prior of the next stage and ii) the evidence for a signal in the searched region.
We now show how we use the $\fstat$-statistic to compute these quantities.

For a candidate, the posterior probability for the phase-evolution parameters $\dparams$ is obtained by marginalizing the full posterior over all possible amplitude parameters $\{\aparams\}$:
\begin{align}
  \probdensity{\dparams |\data} & = \frac{\probdensity{\data |\dparams} \probdensity{\dparams}}{\int_{\searchspace_\candidate} \probdensity{\data |\dparams}\probdensity{\dparams}\diff \dparams}                                                                                                                \\
                                & = \frac{\int_{\{\aparams\}} \probdensity{\data |\aparams,\dparams} \probdensity{\aparams,\dparams} \diff \aparams }{\int_{\searchspace_\candidate} \int_{\{\aparams\}} \probdensity{\data |\aparams,\dparams}\probdensity{\aparams,\dparams} \diff \aparams \diff \dparams} \label{eq:bayesmarginalized}
\end{align}
where ${\searchspace_\candidate}$ is the phase-evolution uncertainty region for that candidate. For detectable signals we will assume that the only nonnegligible contributions of $\probdensity{\data |\aparams,\dparams}$ to the integral above come from $\aparams=\aparams_\mathrm{MLE}$ so we simply set the prior to a delta distribution:
\begin{align}\label{eq:aparamsprior}
  \probdensity{\dparams,\aparams} & = \probdensity{\aparams|\dparams}\probdensity{\dparams}                    \\
                                  & =\delta(\aparams - \aparams_\mathrm{MLE}(\dparams))\probdensity{\dparams}.
\end{align}
This corresponds to a profile-likelihood approach where the amplitude parameters are treated as nuisance parameters.
Using \cref{eq:bayesmarginalized,eq:fstatMLE,eq:likelihoodRatio} with this prior we find
\begin{align}
  \label{eq:thetaPosterior}
  \probdensity{\dparams|\data} \approx \frac{e^{\fstatistic{\dparams}} \probdensity{\dparams}}{\int_{\searchspace_\candidate} e^{\fstatistic{\dparams}} \probdensity{\dparams} \diff \dparams}.
\end{align}
We will come back to the validity of this assumption in \cref{sec:candidate_followup,sec:biases}.

The posterior for semi-coherent searches is obtained as follows. Given data $\data$, first consider a search for $N_\mathrm{seg}$ signals with parameters $\params_\ell$, each constrained to nonoverlapping coherent data segments $\data_\ell$ that span some $T_\mathrm{coh}$.
Because the noise realizations in the different segments $\data_\ell$ are independent random variables, the probability to obtain the set $\dataincoherent$ for signal parameters $\params_0,\dots,\params_\ell \in \Omega$, is given by
\begin{align}
  \probdensity{\dataincoherent|\{\params_\ell\}} = \prod_{\ell = 1}^{N_\mathrm{seg}} \probdensity{\data_\ell|\params_\ell}.
\end{align}
Following Bayes' theorem, the joint posterior distribution for the signal parameters $\{\params_\ell\}$ given the $\{\data_\ell\}$ then reads
\begin{align}\label{eq:parambayesincoh}
  \probdensity{\{\params_\ell\}| \dataincoherent} & = \frac{\probdensity{\{\params_\ell\}}\prod_\ell \probdensity{\data_\ell|\params_\ell}}{\probdensity{\dataincoherent|\Omega \times \cdots \times \Omega}}.
\end{align}

Like before, we treat the $\aparams_\ell$ as nuisance parameters and assume that the only nonnegligible contribution to the evidence integral for each segment comes from the respective $\aparams_{\mathrm{MLE},\ell}(\dparams_\ell)$.
Thus, we apply \crefrange{eq:bayesmarginalized}{eq:thetaPosterior} for each segment to \cref{eq:parambayesincoh} and obtain the marginal posterior
\begin{equation}\label{eq:amplitude_marginal_semicoh}
  \probdensity{\{\dparams_\ell\}|\dataincoherent} = \frac{\probdensity{\{\dparams_\ell\}}\prod_\ell \probdensity{\data_\ell|\dparams_\ell}}{\probdensity{\data|\searchspace_\candidate \times \cdots \times \searchspace_\candidate}}.
\end{equation}
Following \cref{eq:likelihoodRatio},
\begin{align}\label{eq:incohlikelihood}
  \prod_\ell \probdensity{\data_\ell|\dparams_\ell} = \prod_\ell \probdensity{\data_\ell|0}\,e^{\fstatistic_\ell{\dparams_\ell}}.
\end{align}
$\fstat_\ell$ is the coherent $\fstat$-statistic for segment $\ell$.
Note that $\prod_\ell \probdensity{\data_\ell|0} = \probdensity{\data|0}$.

However, following \cref{eq:semcohfstatistic}, semi-coherent approaches target signals with the same phase-evolution parameters in all segments. We can enforce this condition by choosing reference phase-evolution parameters $\dparams$ and setting priors
\begin{equation}
  \probdensity{\{\dparams_\ell\}} \rightarrow \probdensity{\{\dparams_\ell\} |\dparams} = \prod_{\ell} \delta(\dparams_\ell - \dparams).
\end{equation}
We then marginalize \cref{eq:incohlikelihood} over all $\dparams_\ell$ with this prior to yield
\begin{align}
  \MoveEqLeft
  \probdensity{\dataincoherent|\dparams} = \int_{\searchspace_\candidate} \dots \int_{\searchspace_\candidate} \prod_\ell\probdensity{\data_\ell|\dparams_\ell}  \probdensity{\dparams_\ell |\dparams} \diff \dparams_\ell  \\
     & = \probdensity{\data|0}\prod_\ell\int_{\searchspace_\candidate}  e^{\fstatistic_\ell{\dparams_\ell}}\delta(\dparams_\ell - \dparams) \diff \dparams_\ell \\
     & =\probdensity{\data|0} \prod_\ell e^{\fstatistic_\ell{\dparams}} = \probdensity{\data|0}\, e^{N_\mathrm{seg}\semcohfstatistic{\dparams}}, \label{eq:likelihood_semicoh}
\end{align}
where $\bar\fstat$ is the semi-coherent $\fstat$-statistic defined in \cref{eq:semcohfstatistic}.
Again, we find the semi-coherent posterior for the reference phase-evolution parameters $\dparams$ by applying Bayes' theorem to \cref{eq:likelihood_semicoh}:
\begin{align}
  \probdensity{\dparams| \{\data_\ell\}}
   & = \frac{\probdensity{\{\data_\ell\}|\dparams}\probdensity{\dparams}}{\int_{\searchspace_{\candidate}} \probdensity{\{\data_\ell\}|\dparams}\probdensity{\dparams}\diff \dparams}                                                 \\
   & = \frac{e^{N_\mathrm{seg}\semcohfstatistic{\dparams}}\probdensity{\dparams}}{\int_{\searchspace_{\candidate}} e^{N_\mathrm{seg}\semcohfstatistic{\dparams}}\probdensity{\dparams}\diff\dparams}, \label{eq:thetaPosteriorsemcoh}
\end{align}
and rediscover the right side of \cref{eq:thetaPosterior} with the substitution $\fstat \rightarrow N_\mathrm{seg}\bar\fstat$.
The phase-evolution parameter independent noise term $\probdensity{\data|0}$ factors out.

The semi-coherent Bayes' factor for a signal with phase-evolution parameters $\dparams_\signal\in\searchspace_\candidate$ is
\begin{align}
  \mathcal{B}_{\signal/\mathrm{G}} & =  \frac{\probdensity{\{\data_\ell\} | \searchspace_\candidate}}{\probdensity{\{\data_\ell\}| 0}}                                                                             \\
                                   & = \frac{\probdensity{\data | 0}\int_{\searchspace_\candidate}  e^{N_\mathrm{seg}\semcohfstatistic{\dparams}} \probdensity{\dparams} \diff \dparams}{\probdensity{\data | 0}},
\end{align}
which is equal to the semi-coherent evidence for a signal with parameters $\dparams\in\searchspace_\candidate$~\cite{prixSearchMethodLongduration2011a}:
\begin{equation}
  \label{eq:evidence}
  Z(\searchspace_\candidate) = \int_{\searchspace_\candidate}  e^{N_\mathrm{seg}\semcohfstatistic{\dparams}} \probdensity{\dparams} \diff \dparams.
\end{equation}

We note that, in practice, we substitute the semi-coherent $\fstat$-statistic with (half of) the estimated signal power
\begin{equation}\label{eq:signalpower_estimated}
  \hat\rho^2(\dparams) \defeq N_\mathrm{seg} [2\semcohfstatistic{\dparams} - \mathbb{E}_n(2\bar\fstat)] 
\end{equation}
where $\mathbb{E}_n(2\bar\fstat) = 4$, the expectation value of the semi-coherent $2\fstat$-statistic for noise-only data. The substitution factors out in equation \cref{eq:thetaPosteriorsemcoh} and has no effect on the parameter estimation, but removes an $N_\mathrm{seg}$ dependent scaling of the evidence (see \cref{eq:evidence}), which simplifies the comparison of results with different $N_\mathrm{seg}$.

\subsection{Estimating the posteriors and evidences}\label{sec:posts_and_envs}

A direct evaluation of \cref{eq:thetaPosteriorsemcoh} is not possible because the evidence (\cref{eq:evidence}) cannot be expressed in closed form as a function of the data, $\data$. A numerical approach is needed.
One could evaluate
\begin{align}
  \label{eq:thetaPosteriorpropto}
  \probdensity{\dparams|\{\data_\ell\}} \propto e^{N_\mathrm{seg}\semcohfstatistic{\dparams}} \probdensity{\dparams}
\end{align}
over a $\dparams$-template bank as done for grid-based searches, estimate the evidence $Z(\searchspace_\candidate)$ by numerical quadrature~\cite{numericalrecipes}, and approximate the posterior density, e.~g., by interpolation.
However, template banks that resolve the fine structure of the posterior distribution are computationally inefficient, because the majority of statistic evaluations are allocated to regions that contribute little to the evidence integral.

Instead, we use the nested sampling algorithm~\cite{Skilling:2004pqw}. We give a detailed description of nested sampling in \cref{sec:nested_sampling} and refer to~\cite{Ashton:2022grj} for a recent review. In short, the algorithm performs a stochastic evaluation of the evidence integral (\cref{eq:evidence}) using an ensemble of samples randomly drawn from the prior $\probdensity{\dparams}$ -- called live points -- that are iteratively updated toward regions of higher likelihood.

Nested sampling has several advantages for our purpose. On one hand, the algorithm performs a thorough exploration of the parameter space before converging to any found maxima, aiding in the discovery of ``interesting'', signal-like features. 
On the other hand, we apply nested sampling to estimate the evidences and posterior distributions for to up to millions of candidates.
Yet, almost all candidates examined are expected to be noise whereas signal-like features are rare.
Thus, it is crucial to use a sampling algorithm that will converge even if no interesting modes are found.
This is not a trivial property satisfied by nested sampling by virtue of an evidence-based converge criterion (see \cref{eq:convcrit}).

As its main downside, nested sampling may terminate without sampling interesting modes if none of the live points finds them by chance.
The chance for live points to find interesting modes decreases with a higher number of modes, but increases with a higher number of live points.
However, a higher number of live points also linearly increases the runtime of the algorithm (see \cref{eq:convergencerate}).
If significant modes are discovered, all live points are eventually replaced by samples from the most significant mode in later iterations,
and the algorithm is not prone to get ``stuck'' in secondary modes.

The number of possible modes in a search can be estimated from the number of independent templates needed to cover the searched space $\searchspace_\candidate$. A rough estimate is to use the metric of the employed detection statistic and compute the number of unit-mismatch hyper ellipsoids (\cref{eq:metric}) necessary to cover $\searchspace_\candidate$. This quantity is referred to as $N_\star$~\cite{Brady:1997ji,Messenger:2008ta,Ashton:2018ure}, and is often used as a measure of the ``proper'' volume of the search space.

Once converged, nested sampling yields an estimate of the evidence $Z(\searchspace_\candidate)$ and a set of weighted samples $\{w_i, \dparams_i\}$ that can be used to approximate the posterior density (see \cref{eq:sample_weights,eq:estimatedposteriorsamples}). However, the follow-up method we describe in the next section requires a functional form of the posterior density.
We obtain such a form by modeling the posterior probability density with a Gaussian mixture, accounting for the full covariance and multimodality:
\begin{align}\label{eq:gmm}
  \probdensity{\dparams|\{\data_\ell\}} \approx \sum_{k} m_k~\gaussian{\dparams|\mu_k, \Sigma_k}~~{\textrm{with}}~~\sum_k m_k = 1.
\end{align}
This is a weighted sum of multivariate Gaussian distributions $\mathcal{G}$ with means $\mu_k$ and covariance matrices $\Sigma_k$.

We adopt a two-step process to infer suitable model parameters $\{m_k, \mu_k, \Sigma_k\}$ from the weighted samples:
\begin{enumerate}
  \item We use the \texttt{DBSCAN} algorithm~\cite{ester:2004} to identify clusters of samples based on their parameter-space distances.
        We use the semi-coherent phase metric (\cref{eq:metric} and the end of \cref{sec:signalModelAndFstat}) to define a minimum distance between neighboring samples to be part of the same cluster. We also set a minimum occupancy per cluster of at least $\operatorname{dim}(\searchspace) + 1$ (5, in our case) to ensure nondegenerate covariance matrices.
        Samples marked as noise by \texttt{DBSCAN} are dropped.
  \item The weighted samples in each cluster are fitted with a Gaussian mixture model~\cite{GaussianMixtureModels}.
        We do not know a priori the number of modes, so we estimate them from the set of samples.
        We have found that a recursive algorithm that splits the cloud of samples using a model with two Gaussian components, until a one-component model yields a better description, works well.
        We use the Bayesian information criterion~\cite{gelmanBayesianDataAnalysis2014} to quantify the quality of the fit and place thresholds to stop the recursion.
        The thresholds are calibrated with posteriors obtained from simulated signals.
        We define a minimum occupancy per mode. If after a split, a set of points is below this limit, this set is also dropped.
\end{enumerate}
The process prioritizes the modeling of signal-like posteriors.
For any search, the process is validated with a reference population of simulated signals.

\begin{figure}[t]
  \includegraphics[width=\columnwidth]{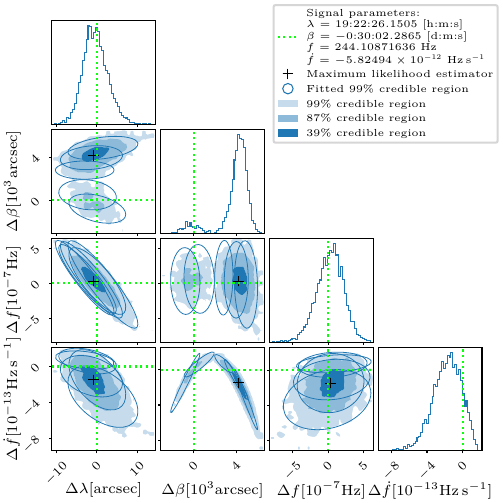}
  \caption{Posterior distribution for a fake signal near the ecliptic equator ($\beta\approx 0$) added to O3a data and using the semi-coherent \fstat-statistic with $T_\mathrm{coh} = 120\,\unit{h}$. The posterior shows a bimodal distribution, split between the hemispheres.
    The plot indicates the $\Delta(C = 0.99)$ region of each mode of the fitted Gaussian mixture model following \cref{eq:credible_regions_gmm}.}
  \label{fig:posterior_bimodal}
\end{figure}
\Cref{fig:posterior_bimodal} shows the Gaussian mixture model for the posterior of a simulated signal.
The signal was placed near the ecliptic equator to produce a characteristic multimodal shape.
The Gaussian mixture model captures the overall shape well with only small discrepancies we consider acceptable.

\subsection{The candidate follow-up}\label{sec:scheme}

We use the inference from one search stage to inform the next one.
Given a posterior distribution $\probdensity^i{\dparams|\{\data_\ell\}}$ for a Stage $i$, we let it serve as the prior for stage $i+1$:
\begin{equation}\label{eq:stagetransition}
  \probdensity^{i+1}{\dparams} = \probdensity^{i}{\dparams|\{\data_\ell\}}.
\end{equation}
We then use nested sampling and \cref{eq:thetaPosteriorsemcoh} to estimate $\probdensity^{i+1}{\dparams|\{\data_{\ell}'\}}$. This conditions the inference at higher coherence times to only consider the most promising parameter-space regions. 
The efficient sampling of the prior in \cref{eq:stagetransition} is done using the inverse-transform method on our Gaussian mixture models (\cref{eq:gmm}) as described in \cref{sec:inversetransform_gmm}.

The process is repeated until the last stage, when the coherence time is equal to the data span.

We note that higher coherence times do not reveal entirely new information.
Repeated application of Bayes' theorem and \cref{eq:stagetransition} may, thus, reuse some information, which incurs compounding biases in the estimated signal parameters.
However, the goal of a search hierarchy is not to produce the most accurate parameter estimation, but to conduct computationally efficient follow-ups.
Any candidates left after the final coherent stage concludes will be subject to more detailed examinations in any case.
Further, if longer coherence times yield sufficient new information, the biases will be negligible.
The extent of this effect on the follow-up presented in \Cref{sec:testingtheframework} is investigated in \Cref{sec:biases}.

Conversely, the initial stochastic stage is initialized with a list of signal candidates marked by seeds $\dparams_s$ and an uncertainty range $\Delta \dparams$ given by the previous, deterministic stage. For each candidate, we choose uninformative priors based on Jeffreys' invariance principle~\cite{gelmanBayesianDataAnalysis2014} that span the respective uncertainty region, $\searchspace_\candidate = \dparams_s \pm \Delta \dparams$.

Jeffreys' invariance principle demands that results from Bayesian inference are invariant under reparametrization of the model.
We obtain such priors by first choosing a parametrization for which the response of the $\fstat$-statistic to a signal does not depend on the signal parameters.
This corresponds to parametrizations for which the $\fstat$-statistic metric is flat.
Using this parametrization, we set uniform priors, and transform them back to our physical parameters.
A detailed derivation can be found in \Cref{sec:initial_priors_details}.

For a search over sky position, frequency and frequency derivative, the $\fstat$-metric is already flat for frequency and frequency derivative.
For the sky position, with a fixed frequency and long data spans, the metric is approximately flat when projected to the ecliptic plane.\footnote{This is why the deterministic searches employ sky position grids that are a uniform (hexagonal) area tiling of the orthogonal projection of each celestial hemisphere on the ecliptic plane.}
The sky position prior for a candidate at a position $\beta,\lambda$ and with frequency $f$ translated from a uniform distribution on the ecliptic plane is accordingly
\begin{equation}\label{eq:skypriorsmaintext}
  \probdensity{\lambda, \beta | f} =
  \frac{2 \pi}{A_\mathrm{ecl}(\searchspace_\mathrm{sky})}\sin(|\beta|)\cos(|\beta|),
\end{equation}
where $A_\mathrm{ecl}(\searchspace_\mathrm{sky})=\pi r^2$ is the area of the sky-uncertainty region on the ecliptic plane (see \Cref{sec:initial_priors_details} for a derivation). This is a circle of radius $r$ on the ecliptic plane, which is inversely proportional to the signal frequency. This circle projects on the celestial sphere to $\searchspace_\mathrm{sky}$, whose position and shape depend on where the circle lies on the ecliptic plane.

\begin{figure}[t]
  \includegraphics[width=.7\columnwidth]{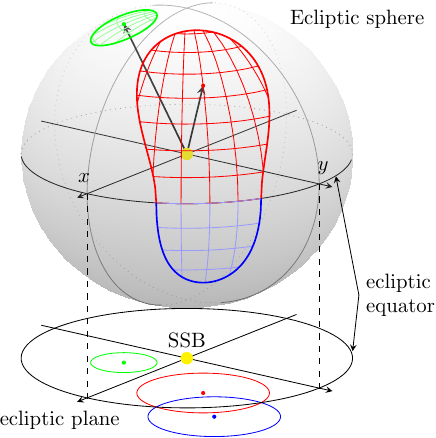}
  \caption{Circles defining two uncertainty regions on the ecliptic plane and their projection to the ecliptic sphere. The size of the uncertainty regions is exaggerated. (Red) Uncertainty region for a candidate that is intersecting the ecliptic equator. (Blue) The extension of the uncertainty region onto the opposing hemisphere. (Green) Uncertainty region of a candidate at higher frequency, which corresponds to a smaller circle on the ecliptic plane.}
  \label{fig:skyparamuncertainties}
\end{figure}
\Cref{fig:skyparamuncertainties} shows the sky parameter uncertainty regions and their projection on the celestial sphere for two examples.
One of these examples depicts a special case that arises when the uncertainty circle on the ecliptic plane intersects the ecliptic equator.
The case corresponds to the $\fstat$-statistic response to signals near the ecliptic equator, which is degenerate in the ecliptic hemispheres.
To account for such cases, we place a second circle with the same radius that intersects the equator in the same locations as the original circle.
Sky points from the opposing hemisphere of the candidate seed are included in the search if their projected ecliptic coordinates lie inside this second circle.

\subsection{Uncertainty on candidates' parameters}

For each candidate, the posterior on the signal parameters defines the \emph{credible region}
\begin{align}
  \label{eq:credible_region}
  \Delta(C) & \defeq \argmin_{\Theta \subset \searchspace_\candidate}\left[\operatorname{Vol}(\Theta): \int_\Theta\probdensity{\dparams | \data} \diff \theta = C \right].
\end{align}
$\Delta(C)$ is the smallest possible subregion of $\searchspace_\candidate$ where the parameters of a putative signal lie with probability $C$.
The definition does not demand $\Delta$ to be connected, thus naturally accounting for multimodal posteriors.

The $C$-credible regions of an unimodal Gaussian with covariance matrix $\Sigma$ are nested ellipsoids with surfaces defined by all $\dparams$ with constant Mahalanobis distances from the mean-vector $\mu$:
\begin{equation}\label{eq:mahalanobis}
  \mathrm{d}^2_\text{Mah}(\dparams) \defeq (\dparams-\mu)\Sigma^{-1}(\dparams-\mu).
\end{equation}
Each distance $\mathrm{d}^2_\text{Mah}$ is associated with the cumulative probability $C$ of the signal parameters laying within that ellipsoid,
given by the CDF of a $\chi^2$-distribution with $\dim(\searchspace)$ degrees of freedom. Inverting this relation, for any $C$, there is some distance
\begin{align}\label{eq:credible_region_gaussian}
  \xi^2(C)  & = \operatorname{F}^{-1}_{\chi^2(\mathrm{df}=\dim(\searchspace))}(C), \\
  \intertext{and, thus,}
  \Delta(C) & = \{\dparams \mid \text{d}^2_\text{Mah}(\dparams) < \xi^2(C) \}.
\end{align}
For a multimodal Gaussian, no closed form expression exists.
Instead, we approximate $\Delta(C)$ from the union of the per-mode credible regions $\Delta_k(C)$:
\begin{gather}\label{eq:credible_regions_gmm}
  \Delta(C) \approx \cup_{k} \Delta_k(C), \\
 \Delta_k(C) = \{\dparams \mid \text{d}^2_\text{Mah}(\dparams, \Sigma_k, \mu_k) < \xi^2(C) \}.
\end{gather}
This approximation overestimates $\Delta(C)$: First, while each $\Delta_k(C)$ contributes $m_k \times C$ to the cumulative probability according to its weight $m_k$, each mode also contributes some residual probability from those points within the credible regions of other modes but not within $\Delta_k(C)$. Thus, the cumulative probability of the union of all $\Delta_k(C)$ is larger than $C$, and so is its volume. Second, the union may include points with lower posterior probability density than some points outside the union.
However, the approximation is sufficient for an order-of-magnitude characterization of credible-region volumes.

Out of all regions $\Theta \subset \searchspace_\candidate$ with $\operatorname{Vol}(\Theta) = \operatorname{Vol}(\Delta(C))$, $\Delta(C)$ \emph{maximizes} the integrated posterior probability and thus must contain the $\{\dparams\}$ with the highest posterior probability density.
Defining
\begin{equation}\label{eq:integrated_posterior}
  X'(\mathcal{P}) \defeq \int_{\{\dparams | \probdensity{\dparams|\{\data_\ell\}} \geq \mathcal{P}\}} \probdensity{\dparams|\{\data_\ell\}} \diff \dparams,
\end{equation}
there is a $\mathcal{P}_C$ such that $X'(\mathcal{P}_C) = C$, and we obtain
\begin{equation}\label{eq:credible_region_via_threshold}
  \Delta(C) = \{\dparams\mid\probdensity{\dparams|\{\data_\ell\}} \geq \mathcal{P}_C\}.
\end{equation}
\Cref{eq:integrated_posterior,eq:credible_region_via_threshold} allow us to associate any parameter point $\dparams$ with a credible level $C_\dparams = X'(\mathcal{P_\dparams})$, where $\mathcal{P_\dparams} = \probdensity{\dparams|\{\data_\ell\}}$.
$C_\dparams$ is the cumulative probability of all parameter points with higher posterior probability density than $\dparams$.

\section{Application on real data}
\label{sec:testingtheframework}
We demonstrate the presented framework in a realistic and demanding setting by applying it to the results of the Einstein@Home search by \textcite{Steltner:2023cfk}
which we refer to henceforth as ``the original search''.

The original search was conducted on public Advanced LIGO data from the third observation run~\cite{KAGRA:2023pio}.
The observation run was split into two subperiods, O3a and O3b.
The first stages of the original search only considered O3a, leaving the O3b data for validation.

The original search targets a $90\%$ detection efficiency for signals with sensitivity depth
\citep{Behnke:2014tma} of $\mathcal{D} = 56\,[\si{1\per\unit{\sqrt{\hertz}}}]$, frequencies $\SI{20}{\hertz} < f_{\signal} < \SI{800}{\hertz}$, first-order spindowns $\SI{-2.6e-9}{\hertz\per\second} < \dot f_{\signal} < \SI{2.6e-10}{\hertz\per\second}$, and sky position parameters uniform all over the sky.

\begin{table}[htbp]
  \centering
  \begin{tabular*}{\linewidth}{@{\extracolsep{\fill}}lrrrrrr}
    \toprule
    & $T_\text{coh}$  & $N_\text{seg}$ & $\Delta f$             & $\Delta\dot{f}$                     & $\frac{r_{\text{sky}}}{d_\text{sky}(f)}$             & $n_\mathrm{live}$ \\ [0.5ex]
    & $\unit{\hour}$  &                & $\unit{\micro\hertz}$  & $10^{-14}\,\unit {\hertz\per\second}$ &   &                   \\ \midrule
    Stage 1     & $120  $         & $37$           & $1000$                 & $11250$                             & $10.0$                       & $1500$            \\ \midrule
    Stage $2$ & $120  $         & $37$           & $50  $                 & $1200 $                             & $2.0$\footnote{Due to an error the Stage 2 sky-uncertainty used for the follow-up is \emph{larger} than necessary. This makes the follow-up more challenging and hence does not invalidate our findings.}                        & $1500$            \\     Stage 3     & $240  $         & $19$           & -                      & -                                   & -                            & $750$             \\
    Stage 4     & $490  $         & $9 $           & -                      & -                                   & -                            & $500$             \\
    Stage 5     & $1100 $         & $4 $           & -                      & -                                   & -                            & $500$             \\
    Stage 6     & $2200 $         & $2 $           & -                      & -                                   & -                            & $500$             \\
    Stage 7     & coh.            & 1              & -                      & -                                   & -                            & $500$             \\ \midrule
    Stage $3^*$ & coh.            & 1              & -                      & -                                   & -                            & $750$             \\ \bottomrule
  \end{tabular*}
  \caption{Per-stage coherence times $T_\text{coh}$, number of coherent data segments $N_\text{seg}$, and live point counts of the search stages. The search regions in frequency, spindown and sky covered by the deterministic follow-up of~\cite{Steltner:2023cfk} are given in the 4$^{\textrm{th}}$-6$^{\textrm{th}}$ columns. These are relevant for Stage 1, where the search regions of the seeds were too large to follow up with our scheme, and for Stage 2 which is the first stage conducted with our framework, and that hence uses those uncertainty regions as the prior support. Stages 3-7 use the posterior of the preceding stage as their prior with the full coherence time hierarchy. We also investigate if Stages 3-7 can be replaced with a single, fully coherent stage, Stage $3^*$, using the Stage 2 posterior as the prior.}
  \label{tab:stagesetups}
\end{table}

\subsection{Setup}

The target detectable signal amplitudes are such that at the output of Stage 0, the signal candidates' detection statistic should be at least at the level of the loudest Gaussian noise candidates.
A stochastic framework like ours could miss such a signal either because the search region is too large -- the number of live points required for the sampler to converge to the signal is computationally not viable -- or because even if there are enough live points for the search region, this is inhabited by very loud disturbances, which the sampler converges to.

The second problem is mitigated by data-preparation and cleaning procedures before any processing is carried out on the data~\cite{Steltner:2021qjy}, and by using line-robust detection statistics to rank the initial search results~\cite{Keitel:2013wga}. Some disturbances do, however, persist and if they resemble an actual astrophysical signal, then one's ability to detect a signal within the same region will be significantly lowered. This is an issue that is common to  stochastic and deterministic searches -- in the latter case typically through top-list saturation effects. This effect is however rare, affecting less than a few percent of the results.

Conversely, the first problem of too large search regions is specific to stochastic methods and hence is very relevant to the characterization of our framework. In particular, since the uncertainty regions around candidates at each stage in the deterministic follow-up shrink as the stages proceed, this problem determines at which stage in the follow-up hierarchy we can begin applying our framework. In principle, we would like to begin the stochastic follow-up directly on the Stage 0 candidates, \ie, with Stage 1.

To determine at which stage we can begin applying our framework we study the recovery of test signals at various stages of the original search. Our detection criterion tests
that the maximum likelihood estimator $\dparams_\mathrm{MLE}$ that the sampler has found is related to the test signal with parameters $\dparams_\mathrm{s}$.
This is enforced by requiring that the power $\hat\rho^2(\dparams_\mathrm{MLE})$ estimated at the maximum likelihood point is close to the \emph{estimated} test signal power $\hat\rho^2(\dparams_\signal)$ (using \cref{eq:signalpower_estimated} for both estimates).
Since the power $\rho^2$ is a monotonic function of the likelihood, if the likelihood sampler has converged to the signal, then it must be that $\hat{\rho}^2(\dparams_\mathrm{MLE}) \geq \hat{\rho}^2(\dparams_\signal)$. On the other hand the $\hat{\rho}^2(\dparams_\mathrm{MLE})$ cannot be too much larger than $\hat{\rho}^2(\dparams_\signal)$, as that would also indicate convergence to some other local maximum. So in all the recovery criterion is:
\begin{equation}\label{eq:necessary_condition}
  \hat\mu_{\min} \leq \hat\mu_\signal \defeq
  \frac{\hat\rho^2(\dparams_\mathrm{s}) - \hat\rho^2(\dparams_\mathrm{MLE})}{\hat\rho^2(\dparams\mathrm{s})}
  \leq 0 + \epsilon.
\end{equation}
We use a tolerance of $\epsilon= 10^{-4}$ for the upper bound to allow for a small offset of $\dparams_\mathrm{MLE}$ from the exact maximum likelihood template due to sampling noise.
We set $\hat\mu_{\min} = -0.125$.
In Gaussian noise
the distance between the $\dparams_\mathrm{MLE}$ and $\dparams_\signal$ is a multivariate Gaussian random variable with covariance matrix $\left[\rho^2 g_{ij}(\dparams_\signal)\right]^{-1}$~\cite{Prix:2006wm}. Following \cref{eq:mahalanobis,eq:credible_region_gaussian,eq:mismatch} the expected distance for the $C$-quantile is
\begin{align}
  \mathrm{d}^2_\text{Mah} =\rho^2 g_{ij}(\dparams_\signal) \Delta\dparams_i \Delta \dparams_j =  \xi^2(C) = \rho_\signal^2 \mu.
\end{align}
The weakest signal from our target population has $\rho^2 \gtrsim 200$. At this amplitude, for $C = 99.995 \%$ then $d^2 \approx 25$ and $\mu \approx 0.125$.

We note that the detection criterion can produce false positives in the early stages of the follow-up because the semi-coherent $\fstat$-statistic response to a signal is generally multimodal (see~\cite{Wette:2015lfa,Allen:2019vcl} and the example posterior shown in \cref{fig:posterior_bimodal}). In this case, unlucky noise fluctuations can elevate secondary modes just enough for the sampler to converge on them, missing the ``correct'' signal parameters.
Another cause of false positives is the presence of disturbances also leading to convergence far from the ``correct'' signal parameters.
The impact of both these factors is reduced substantially at higher coherence times, so that as the coherence time increases at each stage, this empirical detection criterion becomes a more and more accurate proxy.

\begin{figure}[ht]
  \centering
  \includegraphics[width=\columnwidth]{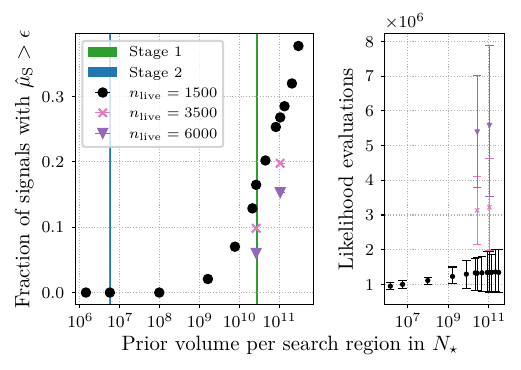}
  \caption{Left: the fraction of missed test signals for search configurations with different prior volumes.
    Right: the average number of per-signal likelihood evaluations for the recovered signals.
    Vertical bars show the 99$^{\mathrm{th}}$ quantile.
    In both plots, vertical colored areas mark configurations listed in \cref{tab:stagesetups}.
    The volumes are given in terms of the number of unit-mismatch hyperellipsoids contained in the search region, $N_\star$.}
  \label{fig:initial_viability}
\end{figure}

We investigate whether we can use our framework for the entire hierarchy of follow-up stages, \ie, starting with Stage 1. So we apply our framework to the test signals seeds as they are produced after Stage 0 in the original search, and to the test signals candidates surviving Stage 1 of the original search. We take as support for the priors the uncertainty regions shown in the first two rows of \cref{tab:stagesetups}.
We use the \texttt{dynesty} sampler~\cite{Speagle:2019ivv} for the nested sampling.
The specific sampler configuration is given in \cref{sec:sampler_config}.
The number of live points for each configuration is given in \cref{tab:stagesetups} or specified explicitly. The recovery results, based on the criterion of \cref{eq:necessary_condition} are shown in \cref{fig:initial_viability}.

At Stage 1, a significant fraction ($10\%\sim 20\%$) of our test signals is lost. A higher number of live points increases the recovery rate, but shows the expected linear scaling in the number of likelihood evaluations (\cref{eq:convergencerate} and~\cite{Ashton:2022grj}).
The number of live points necessary to recover all simulated signals incurs computing costs that are not markedly competitive with the deterministic follow-up.

At Stage 2, with 1500 live points,  we recover all test signals. We take this as our configuration and use our framework for Stages $2 -7$.

After Stage 7 all test signals are still recovered (see \cref{fig:mismatch_coherent} for the distribution of the $\hat\mu$). As a precaution against disturbances, we perform a visual inspection of corner plots of the posterior distributions for each signal.
We do not observe cases where the sampler converged on disturbances.

Through Stages 3-7, we expect that the parameter space volumes to be examined have decreased substantially.
We thus reduce the number of live points as listed in \cref{tab:stagesetups} to improve the computational efficiency. However, the next stages are computationally relatively trivial compared to Stage 2, while recovering all test signals anyway. Thus, we did not optimize the number of live points for later stages. The used numbers of live points here are chosen \textit{ad hoc}.

\subsection{Follow-up of candidates}\label{sec:candidate_followup}

We follow up the $\approx \num{350000}$ candidates produced by Stage 1 of the original follow-up.

\begin{figure}[ht]
  \centering
  \includegraphics[width=\columnwidth]{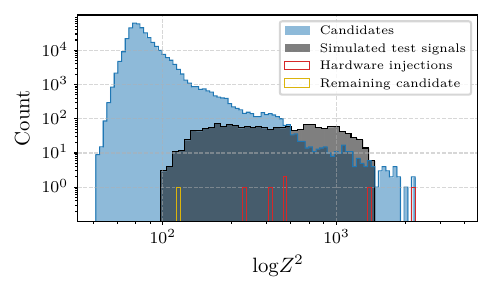}
  \caption{Log-evidences $\log(Z^{2})$ recorded for each candidate and simulated test signal.
    The candidates corresponding to hardware injections and the candidate remaining compatible with the test signal population after Stage 7 are emphasized.}
  \label{fig:logevidencefirststage}
\end{figure}

In \cref{fig:logevidencefirststage} we show the distribution of the evidence from our Stage 2 follow-up.
We can see that the test signals from our target population at this stage do not yet separate from the candidates.

We expect the evidence for a real signal to stay approximately constant as the coherence time of the search increases.
In contrast, disturbances are unlikely to accumulate significance just as a signal would when scrutinized over longer and longer coherence lengths, so we expect that the results of our noise-dominated data will separate from the results of the test signals as the stages progress.

\begin{figure}[ht]
  \centering
  \includegraphics[width=\columnwidth]{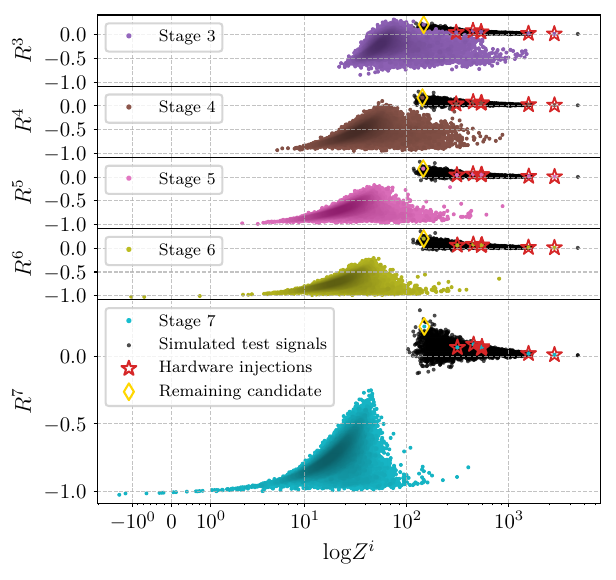}
  \caption{Log-evidences $\log Z^i$ recorded across all follow-ups at stages $i=3,\ldots,7$ and the relative change $R^i$.
    The candidates corresponding to hardware injections and the remaining significant candidate are emphasized. Two hardware injections closely match in $\log Z^i$ and $R^i$ and coincide in the plots at each stage.}
  \label{fig:fullrun_bayes_factor_change_to_first}
\end{figure}

We define the quantity
\begin{equation}\label{eq:R}
  R^{i} \defeq \frac{\log Z^{i} - \log Z^{2^*}}{\log Z^{2}} ~~{\textrm{with}}~~i=3,\ldots,7,
\end{equation}
which characterizes the evolution of the search candidates across the different stages.
\Cref{fig:fullrun_bayes_factor_change_to_first} shows the distributions of the evidences $R^i$ versus $\log{Z^i}$ for all candidates and test signals of each Stage $i=3,\ldots,7$.

As the stages advance the $(\log Z^i,R^i)$ points of the search candidates progressively separate from those of the target signals, and at Stage 7
the bulk of the candidate distribution has markedly separated from the target signal distribution.
At the tail of the bulk (high $\log Z^7$, small $R^7$), a few candidates show log-evidences compatible with our test signals.
However, the decrease in evidence relative to the initial stochastic stage (Stage $2$) of these candidates is not consistent with what is expected from a signal, and hence we do not consider these defensible signal candidates.


Only seven candidates remain compatible with the target distribution of simulated signals.
Six of these candidates are due to hardware injections. These are fake signals added to the data at the hardware level to validate search pipelines~\cite{KAGRA:2023pio}.
The signal parameters of the hardware injections, and their offsets from the respective maximum likelihood estimators are listed in \cref{tab:fullrun_hw_injections}.
The framework recovers the parameters with excellent agreement.

The parameter space examined by the original search contains a seventh hardware injection (ID 11~\cite{KAGRA:2023pio}), which was not recovered by Stage 0 the original search, thus was not in the list of the $\approx \num{350000}$ candidates that we follow up and hence our framework could never recover it. Hardware injection 11 has an amplitude lower than the upper limit level of the original search, and we refer the reader to~\citet{Steltner:2023cfk} for more information.

\begin{table*}[t]
  \centering
  \footnotesize
  \begin{tabular*}{\linewidth}{@{\extracolsep{\fill}}lrrrrrrr}
    \toprule
    $\text{ID}_\text{inj} $ & $f_\text{inj}[\unit{\hertz}]$ &$\dot f_\text{inj}[\unit{\hertz\per\second}]$ & $\alpha_\text{inj} [\text{h:m:s}]$ & $\delta_\text{inj}[\text{deg:m:s}]$ & $\Delta f[\unit{\hertz}]$     &  $\Delta \dot f [\unit{\hertz\per\second}]$& $\Delta_\text{sky}$ [deg:m:s]   \\ [0.5ex] \midrule
    0 & \hwinjzerofreq & \hwinjzerofonedot & \hwinjzeroalpha & \hwinjzerodelta & \hwinjzerodfreq & \hwinjzerodfonedot & \hwinjzerodsky  \\
    2 & \hwinjtwofreq & \hwinjtwofonedot & \hwinjtwoalpha & \hwinjtwodelta & \hwinjtwodfreq & \hwinjtwodfonedot & \hwinjtwodsky  \\
    3 & \hwinjthreefreq & \hwinjthreefonedot & \hwinjthreealpha & \hwinjthreedelta & \hwinjthreedfreq & \hwinjthreedfonedot & \hwinjthreedsky  \\
    5 & \hwinjfivefreq & \hwinjfivefonedot & \hwinjfivealpha & \hwinjfivedelta & \hwinjfivedfreq & \hwinjfivedfonedot & \hwinjfivedsky  \\
    9 & \hwinjninefreq & \hwinjninefonedot & \hwinjninealpha & \hwinjninedelta & \hwinjninedfreq & \hwinjninedfonedot & \hwinjninedsky  \\
    10 & \hwinjtenfreq & \hwinjtenfonedot & \hwinjtenalpha & \hwinjtendelta & \hwinjtendfreq & \hwinjtendfonedot & \hwinjtendsky  \\
    \bottomrule
  \end{tabular*}
  \caption{The frequency, first spindown, right ascension, and declination of the hardware injections, and the distances between these parameters and the maximum likelihood estimators obtained from our follow-up.
    The sky coordinates are in equatorial coordinates.
    The reference time for these values is $\hwinjreftimestageseven$ (GPS time).
    We note the lower offsets reported in the original search using the full O3 data set, \ie, about twice as much data (see Table 3 of~\cite{Steltner:2023cfk}).
  }
  \label{tab:fullrun_hw_injections}
\end{table*}

\begin{figure}[th]
  \includegraphics[width=\columnwidth]{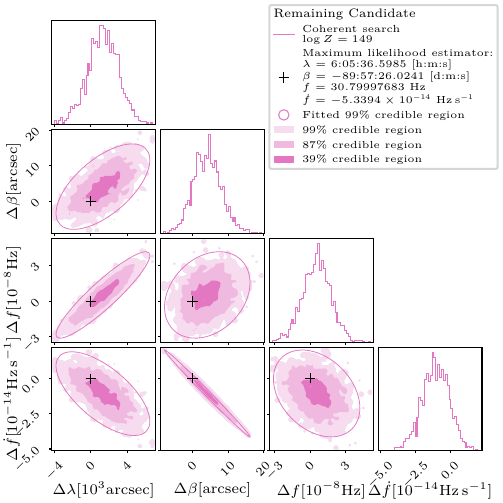}
  \caption{Posterior distribution for the remaining candidate at Stage 7.
    The plot indicates the $\Delta(C = 0.99)$ region of the fitted Gaussian mixture model following \cref{eq:credible_regions_gmm}.}
  \label{fig:posterior_candidate}
\end{figure}

\begin{figure}[th]
  \includegraphics[width=\columnwidth]{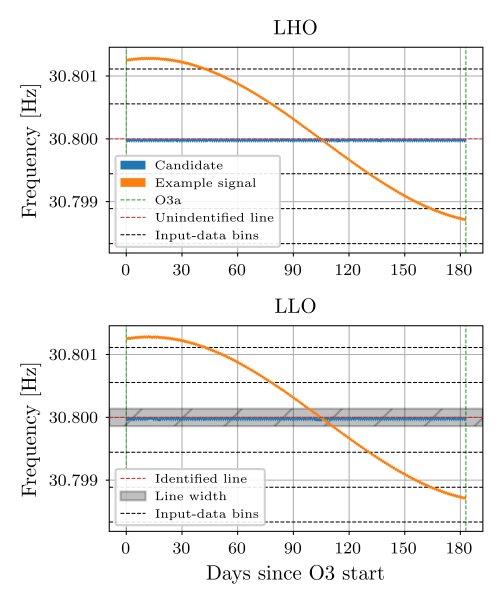}
  \caption{Instantaneous frequency evolution of the remaining candidate in the detector frame of the Hanford (LHO) and Livingston (LLO) observatories and known lines. The signal parameters are that of the maximum likelihood estimator in \cref{fig:posterior_candidate}.
    The same signal but offset in the sky coordinates to ($\lambda = 0, \beta = -1.25$) is shown as an example of a signal modulated by Earth's motion around the sun. The nature of the line in LHO is unknown, whereas the LLO line is confirmed as nonastrophysical.
    The official lines-lists can be found in~\cite{known-lines-list}.
    We mark the input-data bin widths and the line-widths provided in~\cite{known-lines-list} by LIGO.}
  \label{fig:freq_evo_cand}
\end{figure}

After closer inspection of the seventh remaining candidate, we cannot defend it as a significant signal candidate.
\Cref{fig:posterior_candidate} shows the posterior distribution and lists the parameters of the maximum likelihood estimator at the final, coherent stage.
We see that the first derivative of the frequency is close to zero, and that the sky position is at the southern pole of the ecliptic sphere ($\beta\approx-\pi/2$) where the Doppler-modulation due to Earth's motion around the Sun is minimal -- our signal has a meek modulation depth of $\leq 3~\mu$Hz. Astrophysical signals with these parameters are very similar to constant-frequency lines due to coherent, long-lasting disturbances in the data. In the case of the remaining candidate, both detectors feature lines close to/overlapping with the candidates waveform, as shown in \Cref{fig:freq_evo_cand}.
LIGO provides detailed information on lines of known and unknown origin~\citep{known-lines-list}. The line in the Hanford detector is of unknown origin. The line in the Livingston detector was identified as of nonastrophysical origin, however, only after the Einstein@Home was deployed. Thus, this line was not removed in the data preparation phase (see~\cite{Steltner:2021qjy}), resulting in the spurious candidate.
This stresses the importance of data cleaning and noise identification~\cite{Covas:2018oik,Davis:2018yrz,LIGO:2021ppb,Capote:2024rmo}.


The fully coherent stage of the original search on O3a data recorded 12 candidates, including the same six hardware injections, but not the candidate that remains in this follow-up.
The remaining six candidates from the original search were finally rejected by incorporating the O3b data set.
On O3a data alone, the new framework recorded fewer false alarms.

The clean separation of noise candidates and test signals shows that our approximation in deriving the (semi-coherent) $\fstat$-statistic  -- Eq.s~(\ref{eq:aparamsprior},\ref{eq:thetaPosterior},\ref{eq:amplitude_marginal_semicoh}) --  did not negatively affect the effectiveness of the follow-up relative to our population of test signals and the candidates.

\subsection{Search volume compression}
\label{sec:volumeCompression}

\begin{figure}[thbp]
  \centering
  \includegraphics[width=\columnwidth]{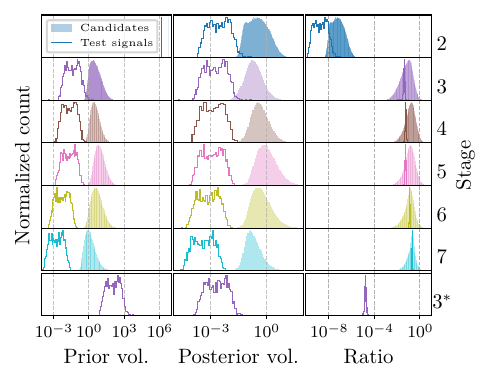}
  \caption{Distributions of the parameter space volumes covered by the prior distributions, the posterior distribution, and their ratio for each stage. The search results for the candidates of the original search are the shaded regions. The solid lines show the results for the target signal population.
    All volumes are given in $N_\star$. All volumes except the prior volumes at Stage $2$ correspond to $\operatorname{Vol}(\hat\Delta(99.9\%))$ of the fitted Gaussian mixture models (see \cref{eq:credible_regions_gmm}).}
  \label{fig:volumes_multistage}
\end{figure}

A further quantity that can be used to characterize the results of the search are the per-candidate prior and posterior volumes at each stage, and the achieved reduction of the volume.
These distributions for each stage are shown in \cref{fig:volumes_multistage}.
Follow-ups of candidates yield significantly larger posterior volumes than those on simulated signals, which translates to larger prior volumes in the next stage.
This difference is explained by the lack of pronounced peaks of the likelihood in follow-ups of candidates.

\subsection{Parameter estimation}\label{sec:biases}

If the inferred Gaussian mixture model posterior density $\probdensity{\dparams | \{\data_\ell\}}$ estimated from data containing a signal with parameters $\dparams_\signal$ accurately models the parameter uncertainties,
the posterior must be consistent with $\dparams_\signal$:
\begin{equation}\label{eq:ppplot_equation}
  \prob{\dparams_\signal \in \Delta(C)} = C.
\end{equation}
We can evaluate the accuracy of our posteriors by estimating $\prob{\dparams_\signal \in \Delta(C)}$ using the results from the reference signal population.
This procedure leads to the ``pp-plots'' (probability-probability plots) that are often used to validate results from Bayesian parameter estimation.

Let us fix a stage. For each of the $N_\mathrm{signal}$ test signals with parameters $\dparams_\signal$, we take the Gaussian mixture model posterior $\probdensity{\dparams| \{\data_\ell\}}$ and estimate $X'(\mathcal{P}_\signal)$ of \cref{eq:integrated_posterior} as
\begin{equation}
  \label{eq:XprimePsig}
  X'(\mathcal{P}_\signal) \approx \frac{\sum_{j=1}^N \boldone_{\{\dparams \mid \probdensity{\dparams| \{\data_\ell\}}> \mathcal{P}_\signal\}}(\dparams_j)}{N},
\end{equation}
with $\mathcal{P}_\signal= \probdensity{\dparams_\signal| \{\data_\ell\}}$, $N = \num{100000}$ being the total number of $\dparams_j$ samples drawn from the posterior and $\boldone_X(x)$ the indicator function
\begin{equation}
  \label{eq:indicator}
  \boldone_X(x) \equiv
  \begin{cases}
    1 & \text{if $x \in X$} \\
    0 & \text{otherwise.}
  \end{cases}
\end{equation}
$X'(\mathcal{P}_\signal) $ of \cref{eq:XprimePsig} is the probability of parameter-space volumes with posterior density higher than that of the signal.
We then count the fraction of test signals for which $X'(\mathcal{P}_\signal) < C$ and take it as an estimate of $\prob{\dparams \in \Delta(C)}$:
\begin{equation}
  \prob{\dparams_\signal \in \Delta(C)} \approx \frac{\sum_\signal \boldone_{\{\mathcal{P} \mid X'(\mathcal{P}) < C\}}(\mathcal{P}_\signal)}{N_\mathrm{signal}}.
\end{equation}

\begin{figure}[ht]
  \centering
  \includegraphics[width=\columnwidth]{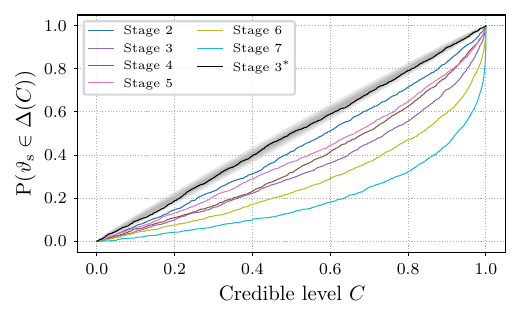}
  \caption{Probability-probability plot showing the test signal recovery for all stochastic follow-up stages. The $x$-axis plots the credible level interval $[0,1)$. The $y$-axis plots the (frequentist) probability of recovering test signals within a credible level $C$, $\prob{\dparams_\mathrm{s} \in \Delta(C)}$.
  Plotted in gray are the 1, 2 and $3\sigma$ confidence regions expected for unbiased parameter estimation given the number of test signals.}
  \label{fig:recovery_multistage}
\end{figure}

\Cref{fig:recovery_multistage} shows $\prob{\dparams \in \Delta(C)}$ as a function of $C$ for all stochastic search stages. Systematic biases in the parameter estimation would lead to deviations from \cref{eq:ppplot_equation}.

Already at Stage $2$ more test signals than expected are found in low posterior probability regions, corresponding to $\prob{\dparams \in \Delta(C)} < C$. This shows that the posterior is slightly biased toward being narrower than it actually is, and indicates that for this stage the underlying parameter uncertainties are not captured perfectly.
Two approximations made by our framework are the likely causes. The first approximation was made in deriving the $\fstat$-statistic based version of Bayes' theorem in Eq.s~(\ref{eq:aparamsprior},\ref{eq:thetaPosterior}\ref{eq:amplitude_marginal_semicoh}), where we assumed that only the most likely amplitude parameters $\aparams_\mathrm{MLE}$ contribute to the evidence integral. Instead, if small offsets in the phase-evolution parameters can be absorbed by small offsets in the amplitude parameters, the posterior distributions would be wider than our results.
The second approximation is the fitted Gaussian mixture model, which may not be able to fit the posterior distributions perfectly.

For Stages $3-7$, the bias increases further.
This is the expected behavior for repeated application of Bayes' theorem using the same information, as anticipated in \cref{sec:scheme}.

We compare these results to a modified coherence time hierarchy of the follow-up after Stage $2$:
Instead of applying Stages $3-7$, we immediately transition to full coherence (Stage $3^*$ in \cref{tab:stagesetups}).
Because of the steep increase in coherence time, more information is used for the inference at Stage $3^*$, and the effect of the compounding bias is expected to lessen.

On the other hand, the steeper increase of the coherence time also means that the sampler has to explore a prior volume resolving more independent templates (\cref{fig:volumes_multistage}), which could result in the loss of signals similar to our examination of Stage 1.
We evaluate \cref{eq:necessary_condition}  and find that no signals are lost.

\Cref{fig:recovery_multistage} shows the $(C,\prob{\dparams_\signal \in \Delta(C)})$-graph for Stage $3^*$ alongside the other stages. The graph agrees well with the expected shape for unbiased parameter estimation.
This result also indicates that at full coherence, the approximation made in deriving Bayes' theorem for the $\fstat$-statistic does not appear to negatively affect the parameter estimation.

Thus, considering only biases in the parameter estimation, a two-stage nondeterministic setup performs better than the full stage hierarchy.

\begin{figure}[ht]
  \centering
  \includegraphics[width=\columnwidth]{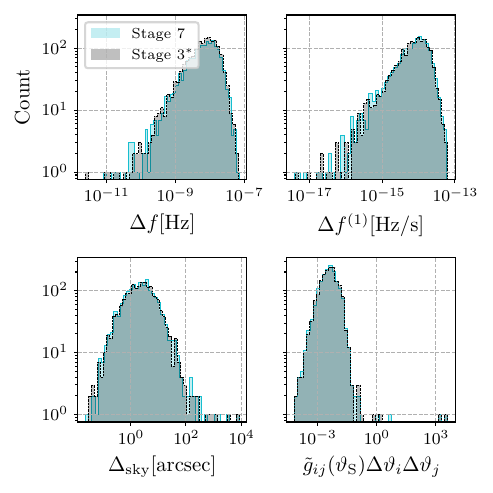}
  \caption{Parameter-space distances between the signal parameters $\dparams_\signal$ and the maximum likelihood estimators $\dparams_\mathrm{MLE}$ at the coherent search stages (Stage 7 and $3^*$) of each test signal, and the calculated  metric-based distance following \cref{eq:mismatch,eq:metric,eq:phasemetric}.
  }
  \label{fig:mismatch_coherent}
\end{figure}

As a further check, we also investigate the frequentist uncertainties of the maximum likelihood estimators by considering the distribution of the parameter-space distances between the maximum likelihood estimators and the test signal parameters.
The distributions are shown in \cref{fig:mismatch_coherent}.
For $f$ and $\dot f$, the uncertainties are $\Delta f \lesssim 10^{-7}\,\unit{\hertz}$ and $\Delta \dot f \lesssim 10^{-13}\,\unit{\hertz\per\second}$, respectively.
For the sky position uncertainty, we find $\Delta_\mathrm{sky} \lesssim 100~\mathrm{arcseconds}$ for almost all signals, and it is centered at a few arcseconds, consistent with the investigated frequency range and an effective ``aperture" of 1 AU.

We note five outliers.
The outliers correspond to signals with sky position parameters very close to the ecliptic equator, and due to noise, the maximum likelihood estimators lie on the ``wrong'' ecliptic hemisphere.
These same signals also produce outliers when considering the metric-based distance measure (lower right plot in \cref{fig:mismatch_coherent}): The metric does not account for multi-modality, and, thus, significantly overestimates the mismatch when the $\fstat$-statistic has multiple, nearby modes.
\textcite{Wette:2015lfa} provides a different parametrization for the metric that accounts for the specific multi-modality encountered. In that parametrization, the calculated metric-mismatch of these outliers would be significantly reduced.

\subsection{Computational cost}\label{sec:compcost}

\begin{figure}[t]
  \centering
  \includegraphics[width=\columnwidth]{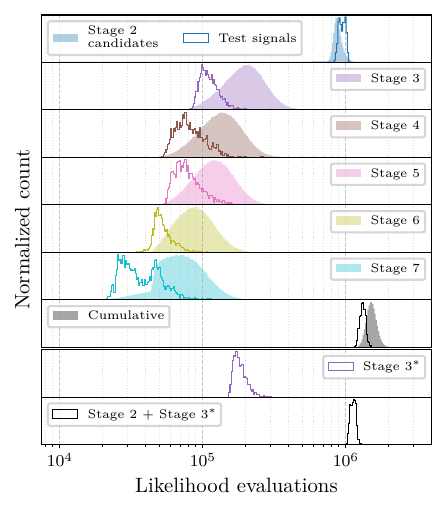}
  \caption{Distribution of the number of likelihood evaluations performed per simulated signal or candidate at each stage, and cumulative for each simulated signal or candidate across all stages.}
  \label{fig:likevals_multistage}
\end{figure}

\Cref{fig:likevals_multistage} shows the number of likelihood evaluations per stage for candidates and test signals.
The cost of the initial of Stage 2 on the candidates is comparable with that on simulated test signals.
The slight difference -- with the searches on noise taking slightly less time than those on simulated signals -- is due to the presence of pronounced peaks in the signal likelihoods, which result in larger evidences that take longer to integrate (see \cref{eq:convcrit}).

Conversely, in the subsequent stages, the noise candidates take longer to follow up than the signals. The reason is that the noise candidates typically have wider posteriors compared to signal candidates,
thereby larger prior volumes to sample in the next stage (see \cref{fig:volumes_multistage}).

Stages $3-7$ require significantly fewer likelihood evaluations per candidate compared to Stage 2, for both the candidates and for the test signals.
The reduction in computational cost that ensues makes it viable to apply Stages $3-7$ to {\it{all}} candidates without any intermediate candidate vetoing, and this incurs
for the full hierarchy only $1.5$ times the computational cost of Stage $2$ alone. In contrast, deterministic follow-up schemes require vetoing candidates at all intermediate stages, which requires nontrivial safety studies.

For the simulated test signals, the immediate coherent follow-up in Stage $3^*$ requires approximately twice as many likelihood evaluations as the semi-coherent Stage $3$, caused by the higher number of waveforms resolved compared to Stage 3.
However, because of the smaller number of stages, the total computational cost is smaller than the full hierarchy.
We refrain from reapplying Stage $3^*$ to the search candidates because of the computational cost of a repeated analysis.

To provide context for these results, we give an order-of-magnitude comparison to the signal template grids used by the original search.
These numbers are estimates from the description in~\cite{Steltner:2023cfk}.

In the original search, the deterministic Stages $2-7$ examine $\approx 10^{9}$ templates per candidate and stage.
Cumulatively, and accounting for candidate rejections after each stage, a total of $4\times 10^{14}$ templates are examined.

Assuming $10^{-2}\,\unit{s}$ per likelihood evaluation and neglecting overhead~\cite{Covas:2024pam}, on a computer cluster with $10^4$ CPUs, this yields a runtime of $\mathcal{O}(\mathrm{days})$ for our follow-up, which closely matches the actual runtime on the ATLAS computing cluster.
For a deterministic search and with the same cost per likelihood evaluation, this would yield $> \mathcal{O}(years)$ for the original search with candidate rejection.
However, we emphasize that searches with grid-based template banks, such as the original search, use optimized algorithms to compute the \fstat-statistic that cache certain intermediate results and (significantly) decrease their computational cost.
The exact speedup has complex dependencies on the exact setup and available hardware.
With these optimizations and accounting for stage setup and accounting for postprocessing, Stages 2-7 of the original follow-up in practice required $\mathcal{O}(\mathrm{week})$ in combined human and computing time \emph{per stage}.

\section{Conclusion and Discussion}\label{sec:conclusion}

We presented a novel Bayesian framework for large-scale follow-ups of candidates of continuous wave searches.
The framework implements a novel stage transition scheme that uses the posterior distribution of one stage as the prior distribution of a new search stage with a higher coherence time.

We present results of a real-world application of  this framework to results of the O3 data Einstein@Home all-sky search for continuous waves emitted by isolated neutron stars~\citep{Steltner:2023cfk}.

We are able to apply the framework to the candidates produced by the Stage 1 follow-up of that search, covering search regions with per-candidate volumes of $N_\star \lesssim 10^8$. This is consistent with recent results~\cite{Covas:2024pam} and exceeds the volumes that were previously thought to be the limit of Monte Carlo based frameworks by several orders of magnitude.

We show a significant reduction in the computational cost of subsequent stages and achieve the coherent follow-up of \emph{all} $\approx \num{350000}$ candidates.
To our knowledge, no follow-up \emph{without vetoing} between stages on this scale has been reported.
Our final search results agree with the original search.

We emphasize the significant reduction of the manual and computational overhead achieved by this approach compared to traditional deterministic follow-up frameworks.
Once a viable sampler configuration for an initial target search region volume is found, the full follow-up hierarchy is performed on every candidate to the end.
Reference to a target signal population is necessary only for the initial configuration and the final thresholding.

The framework provides a significant simplification of the original follow-up procedure. Still, the necessity of one traditional, grid-based follow-up stage remains for the first stage, due to the large uncertainties in signal parameters in the candidates from that stage.
Pushing toward even larger initial volumes is necessary to replace existing follow-up frameworks entirely.

In order to compare with the deterministic follow-ups~\citep{Steltner:2023cfk}, we use the coherence time hierarchy of that search.
However, we show that our framework performs better with a two-stage setup compared to the full coherence time hierarchy when applied to the test signals: The computational cost is reduced and the systematic biases of the parameter estimation are eliminated. This is relevant looking ahead at a follow-up that produces interesting candidates: one of the next steps would then be to search for an electromagnetic counterpart signal informed by our parameter estimates.

Our follow-up only considers O3a data, approximately half of the available time-baseline of the third LIGO observation period~\cite{KAGRA:2023pio}, and about one fifth of the planned  time-baseline of the fourth LIGO-Virgo-KAGRA observation period~\cite{ligo-o4-observing-plan-current}.
We consider it likely that a two-stage setup is not viable for such a long time-baseline.
If intermediate stages remain necessary, the \texttt{PYFSTAT} framework and its ``offsprings'' offer optimization schemes for the construction of the coherence time hierarchy of stochastic searches~\cite{Ashton:2018jfp,Tenorio:2021njf,Mirasola:2024lcq}.
The optimization is based on the \emph{expected} prior volumes of the \emph{population} of candidates at each stage. In our framework, the search regions defined by \cref{eq:credible_regions_gmm} would enable a per-candidate optimization based on the actually \emph{observed} volumes -- once we implement these schemes.

All-sky searches for continuous waves from neutron stars in binary systems are plagued by the computational cost steeply rising with increasing search coherence length. The development of frameworks like this opens prospects for carrying out extensive follow-ups in these challenging higher dimensionality parameter spaces~\cite{Covas:2024pam,KAGRA:2021vkt}.

\section*{Acknowledgments}
This project made use of \texttt{LALSUITE}~\cite{lalsuite}, \texttt{SWIGLAL}~\cite{Wette:2020air}, and the \texttt{scikit-learn} Python package~\cite{Pedregosa:2011ork}. The work utilized the ATLAS computing cluster at the MPI for Gravitational Physics Hannover. This project has received funding from the European Union's Horizon 2020 research and innovation program under the Marie Sklodowska-Curie grant agreement number 101029058.

\appendix

\section{Nested sampling}\label{sec:nested_sampling}
Nested sampling is a Monte Carlo algorithm designed to estimate evidences~\cite{Skilling:2004pqw}.
The method is based on a reformulation of the evidence integral.
The prior probability over parameter space points $\{\dparams\}$ such that the likelihood $\probdensity{\{\data_\ell\}|\dparams}$ exceeds a certain value $\likelihood$
\begin{equation}\label{eq:priorspace}
  X(\likelihood) \defeq \int_{\{\dparams \in \searchspace_\candidate | \probdensity{\{\data_\ell\}|\dparams} > \likelihood\}} \probdensity{\dparams} \diff \dparams,
\end{equation}
allows us to express \cref{eq:evidence} as
\begin{equation}\label{eq:nestedintegral}
  Z(\searchspace_\candidate) = \int_0^{\infty} X(\likelihood) \diff \likelihood = \int_{0}^{1} \likelihood(X) \diff X.
\end{equation}

Nested sampling stochastically estimates the $X_i$ and associated $\likelihood(X_i)$ through an iterative process.
A set of $n_\mathrm{live}$ parameter-tuples, called live points, is sampled from the prior distribution $\probdensity{\dparams}$. At iteration $i$ the live point with the smallest likelihood defines the threshold $\likelihood^t_{i}$.
New samples are proposed until one exceeds $\likelihood^t_{i}$. The point is accepted as a live point and the lowest likelihood point is removed. This generates a new set of $n_\mathrm{live}$ live points for iteration $i+1$. A new threshold $\likelihood^t_{i+1}> \likelihood^t_{i}$ is defined, and the process repeats itself.
The prior probability $X(\likelihood^t_i)$ is expected to decrease exponentially (see \cref{eq:remainingvolume}). The  $X_i$ is the prior probability of {\it{all}} parameter space points with likelihood exceeding $\likelihood^t_{i}$. The parameter space volumes defined by these points are nested, in the sense that all points in the volume at iteration $i+1$ are also in the volume of iteration $i$.

The ``shrinkage'' of  $X_i$ relative to $X_{i-1}$ is independent of $i$ and follows a Beta-distribution:
\begin{equation}
  Y:=  \frac{X_i}{X_{i-1}}\sim \operatorname{Beta}(1, n_\mathrm{live}), \quad X_0 = 1.
\end{equation}
The expected value of $\log Y$ is~\cite{Skilling:2004pqw}
\begin{equation}
  \mathbb{E}[\log Y]=-1 / n_\mathrm{live}
\end{equation}
and hence
\begin{equation}\label{eq:remainingvolume}
  \mathbb{E}[\log X_k] =  \mathbb{E}\left[\log \frac{X_k}{X_{0}}\right] \approx -k / n_\mathrm{live}.
\end{equation}
This last equation allows to estimate the $X_i$ values in
\begin{equation}\label{eq:evidenceestimate}
  Z_k \approx \sum_{i}^{k} \frac{1}{2} (X_{i-1} - X_{i+1}) \likelihood^t_i.
\end{equation}
The expected number of iterations necessary such that the remaining probability $X_k$ reaches some reference $X^*$ scales as
\begin{equation}\label{eq:convergencerate}
  \mathcal{O}(n_\mathrm{iter}) \approx n_\mathrm{live} \log X^*,
\end{equation}
\ie, linear in the number of live points but logarithmically in $X^*$.

At each iteration, we want to find a new sample with likelihood $>\likelihood^t_i$.
The efficiency of this process depends on the number of trials necessary until a sample is accepted.
The simplest way to implement this is to sample the initial prior directly.
However, the smaller the prior probability of the parameter space that meets this condition ($X_i$), the greater the number of trials necessary for a randomly sampled point to have likelihood $>\likelihood^t_i$.
Since $X_i$ shrinks exponentially with each iteration, drawing samples from the initial prior volume becomes exponentially less efficient.
In short, the method still allocates most likelihood evaluations to uninteresting parameter space regions.

Implementations of nested sampling offer different algorithms to solve this problem.
An in-depth discussion of these algorithms is beyond the scope of this paper.
We refer the interested reader to~\cite{Ashton:2022grj,Speagle:2019ivv}.

We use a random walk, following the well-established Metropolis-Hastings algorithm~\cite{Hastings:1970}.
Starting at a randomly chosen live point, a series of random steps (each with a randomly chosen direction) are taken.
Steps are accepted only if the stepped-upon point exceeds $\likelihood_i^t$.
After some number of steps $n_\mathrm{steps}$ the current position can be taken as a new, independently sampled live point.
The computational cost per iteration is constant, and, following \cref{eq:convergencerate}, the overall cost to reach some reference $\log X^*$ remains linear.

Whatever method is chosen for finding the new live point, they all involve sampling the prior. The most efficient way to do this is the inverse transform method.
The method takes a function $g$ for which
\begin{equation}\label{eq:probdensitymapping}
  \probdensity{g(U) = \dparams} = \probdensity{\dparams},
\end{equation}
where $U$ is a uniform random variable. Then, $\dparams = g(u)$ is a sample from the prior.
The function $g$ is the inverse cumulative distribution function of the prior -- if it exists in closed form.

After $k$ iterations, the remaining evidence is (\cref{eq:evidenceestimate,eq:nestedintegral})
\begin{equation}\label{eq:remaining_evidence}
  \Delta Z_k \defeq Z(\searchspace_\candidate) - Z_k =  \int_{0}^{X_k} \likelihood(X) \diff X.
\end{equation}
An upper bound on $\Delta Z_k$ can be set assuming that all points within the remaining volume have the same likelihood:
\begin{equation}\label{eq:remaining_evidence_bound}
  \Delta Z_k \leq \max_\dparams\probdensity{\{\data_\ell\}|\dparams} X_k.
\end{equation}
We can estimate $\max_\dparams \probdensity{\{\data_\ell\}|\dparams}$ with the highest likelihood of the current live points,  $\likelihood_{{\max}, k}$, and use \cref{eq:remaining_evidence_bound} as a stopping criterion for the evidence integration:
\begin{equation}\label{eq:convcrit}
  \frac{\Delta Z_\mathrm{k}}{Z_k} \lesssim \frac{\likelihood_{\max,k} X_k}{Z_k} < \delta^t,
\end{equation}
where $\delta^t$ is some chosen threshold. For low $k$, the estimate $\likelihood_{{\max}, k}$ has a large uncertainty. However, for low $k$, \cref{eq:convcrit} is dominated by the still large $X_k$ and small $Z_k$.
Once $\likelihood_{{\max}, k}$ and $Z_k$ start to converge, \cref{eq:convcrit} is dominated by $X_k$ which continues to decrease exponentially.

Once converged, nested sampling yields the integrated evidence, $Z\approx Z_{n_\mathrm{iter}}$, and a set $\{\dparams_i\}$ with $i=1,\ldots,n_{iter}$, where $\{\dparams_i\}$ are the live points that were \emph{replaced} by a new sample at each iteration $i$.
The local density of the points $\dparams_i$, weighted by their relative contribution to the evidence integral
\begin{equation}\label{eq:sample_weights}
  w_i = \frac{1}{Z}\left[\frac{1}{2}(X_{i-1} - X_{i+1})\likelihood_i\right],
\end{equation}
approximates the (local) posterior probability:
\begin{equation}\label{eq:estimatedposteriorsamples}
  \sum_{i} w_i\,\boldone_{\Delta\dparams}(\dparams_i) \approx \int_{\Delta\dparams}\probdensity{\{\dparams_\ell\}|\data}\diff \dparams,
\end{equation}
with $\boldone_{\mathcal{Y}}(y) = 1$ if $y \in \mathcal{Y}$, and $0$ otherwise.
In \cref{sec:scheme}, we discuss how we construct a functional approximation of the posterior density function from the weighted $\dparams_i$.
\section{Priors at the initial stage}\label{sec:initial_priors_details}

The initial search stage of the stochastic framework uses uninformative priors based on Jeffreys' invariance principle.
The principle demands Bayesian inference to be invariant under reparametrization of the likelihood,
which can be achieved by setting \cite{gelmanBayesianDataAnalysis2014}
\begin{equation}
  \probdensity{\dparams} \propto \sqrt{\det \Gamma_{ij}(\dparams)}.
\end{equation}
$\Gamma_{ij}(\dparams)$ is the Fisher information matrix.
If $\dparams$ is a parametrization for which $\Gamma_{ij}(\dparams)$ is constant, $\probdensity{\dparams}$ is a uniform distribution.

We want to find such priors for the IT1-signal model we consider, typically parameterized by frequency $f$, spindown $\dot f$, and the sky position in equatorial coordinates $\alpha$ and $\delta$ or ecliptic coordinates $\lambda$ and $\beta$.

The Fisher information matrix is proportional to the $\fstat$-statistic metric \citep{Prix:2006wm}, which is constant in the frequency and spin down components, but generally not in the sky. The $f$ and $\dot f$ metric components are constant: they do not depend on the parameter space point. The metric $\alpha$ and $\delta$ components on the other hand do depend on the parameter-space point, in particular they always depend on $f$. They also depend on sky position, but for observation times much longer than days, the metric is approximately constant in the sky when projected to the ecliptic plane \cite{Wette:2015lfa}. We neglect the correlations between $\dot f$ and sky.

Since the candidates' uncertainty in frequency is very small for each candidate we can consider the frequency as fixed. We will hence take uniform priors for $f$, $\dot f$ and $(x, y)$, which are the orthogonally projected sky points on the ecliptic plane:

We start with a uniform distribution on the unit circle ($r \leq 1$) describing the ecliptic plane, parameterized in polar coordinates
\begin{equation}
  (x, y) = (r \cos(\varphi), r \sin(\varphi)),\quad 0 \leq \phi < 2 \pi.
\end{equation}
In this parametrization, a uniform distribution is given by
\begin{align}
    \probdensity{\varphi} = \frac{1}{2 \pi}, \qquad
    \probdensity{r} = 2 r.
\end{align}
The validity is easily checked by integrating over circles with different radii. 

The projection of a point $(r \cos(\varphi), r \sin(\varphi))$ on the ecliptic plane to the surface of the unit sphere (in spherical coordinates) is achieved by
\begin{align}
    \lambda = \varphi \qquad
    |\beta| = \arcsin\left(\sqrt{(1 - r^2)}\right),
\end{align}
where there are two solutions for $\beta$, one for each hemisphere.
The coordinates $(\lambda, \beta)$ are the ecliptic longitude and latitude, respectively.

Evidently, $\prob{\lambda} = \prob{\varphi}$.
The distribution for $\beta$ is given by the Jacobian of the transformation \cite{gelmanBayesianDataAnalysis2014}.
Considering only  positive values of $\beta$,
\begin{align}
    \probdensity{\beta_+}
     & = \probdensity{r(\beta_+)} \cdot \left|\frac{\partial r(\beta_+)}{\partial \beta_+}\right|\\
     & = \probdensity{r(\beta_+)} \cdot \left|\frac{\partial}{\partial \beta_+}\left(\sqrt{1 - \sin^2(\beta_+)}\right)\right| \nonumber \\
     & = 2r(\beta_+) \frac{\sin(\beta_+)\cos(\beta_+)}{r(\beta_+)} \nonumber \\
     & = 2 \sin(\beta_+)\cos(\beta_+). \nonumber
\end{align}
Negative $\beta$ values yield the same distribution. The symmetry implies
\begin{align}
     \probdensity{\beta} &= \probdensity{\operatorname{sign}(\beta)}\probdensity{\operatorname{sign}(\beta)\,\beta} \\
     &= \frac{1}{2} \cdot 2 \sin(|\beta|)\cos(|\beta|),
\end{align}
and the joint probability reads
\begin{align}
     \probdensity{\lambda, \beta} &= \probdensity{\lambda} \cdot \probdensity{\beta} = \frac{1}{2 \pi} \sin(|\beta|)\cos(|\beta|).
 \end{align}

If the distribution is constrained to some subregion $\mathcal{R}_\text{sky}$, the corresponding probability density is
\begin{equation}\label{eq:skypriors}
  \probdensity{\lambda, \beta | \searchspace_\mathrm{sky}} = \begin{cases}
    \frac{2\pi}{A_\mathrm{ecl}(\mathcal{R}_\text{sky})} \probdensity{\lambda, \beta} & \text{if } (\lambda, \beta) \in \mathcal{R}_\text{sky}, \\
    0                                                                                     & \text{else},
  \end{cases}
\end{equation}
where $A_\mathrm{ecl}(\mathcal R_\text{sky})$ is the projected area spanned by $\mathcal R_\text{sky}$ on the ecliptic plane, summed over both hemispheres.
This is simply a consequence of 
\begin{equation}
  \probdensity{x_\mathrm{ecl}(\lambda,\beta),y_\mathrm{ecl}(\lambda,\beta) | \searchspace_\mathrm{sky}} = \frac{1}{A_\mathrm{ecl}(\mathcal R_\text{sky})}
\end{equation}
being a uniform distribution by construction.

The CDF for $\beta$ is  
\begingroup
\allowdisplaybreaks
\begin{align}
    u(\beta):= \operatorname{CDF}(\beta) & = \int_{-\pi/2}^\beta  \sin(|\beta'|)\cos(|\beta'|) \diff{\beta'}    \\   
                      & =  \int_{-\pi/2}^\beta \operatorname{sign}(\beta') \frac{1}{2} \sin(2\beta')\diff{\beta'} \nonumber\\                       
                      & = - \frac{1}{2}\int_{-\pi/2}^0 \sin(2\beta') \diff{\beta'} \nonumber \\ 
                      &\qquad + \frac{1}{2}\operatorname{sign}(\beta) \int_{0}^\beta \sin(2\beta') \diff{\beta'} \nonumber \\
                      & = \frac{1}{2} + \frac{1}{2}\operatorname{sign}(\beta) \left[-\cos^2(\beta')\right]_0^\beta                  \nonumber  \\
                      & = \frac{1}{2} + \frac{1}{2}\operatorname{sign}(\beta) \sin^2(\beta), 
\end{align}
\endgroup
which can be inverted in closed form:
\begin{equation}\label{eq:inversebetacdf}
   \beta(u) = \arcsin\left(\sqrt{|2u - 1|}\right) \operatorname{sign}(2 u - 1).
\end{equation}
This enables sampling $\beta$ from a uniform distribution. 

If $\beta$ is constrained to a sub range, $\beta \in [\beta_{\min}, \beta_{\max}]$, similar forms can be obtained by renormalizing the $u(\beta)$.
However, we note that the constraint applied in \cref{eq:skypriors} yields a joint distribution where $\lambda$ and $\beta$ are not necessarily independent.
In such cases, \eg when $\searchspace_\mathrm{sky}$ describes a circle on the ecliptic plane, \cref{eq:inversebetacdf} can still be used by discarding any samples beyond $\searchspace_\mathrm{sky}$.

\section{The inverse transform method for Gaussian mixture models}\label{sec:inversetransform_gmm}

In our scheme, we approximate the posterior distribution of the phase-evolution parameters $\dparams$ with a Gaussian mixture model:
\begin{align}
  \probdensity{\dparams|\{\data_\ell\}} \approx \sum_{k} m_k~\gaussian{\dparams|\mu_k, \Sigma_k}~~{\textrm{with}}~~\sum_k m_k = 1,
\end{align}
where $\mathcal{G}(\mu_k, \Sigma_k)$ are multivariate Gaussian distributions with means $\mu_k$ and covariance matrices $\Sigma_k$, weighted by weights $m_k$.
Here, we describe how such a prior can be sampled using the inverse transform method (\cref{eq:probdensitymapping}). 

For a single-mode, multivariate Gaussian, the transformation $g$ from uniform random variables $u\sim{\mathcal{U}}$ to our-phase evolution $\dparams$ variables is
\begin{equation}
  \dparams^j = g^j(u) = [L^{-1}]^{j}_{\,l} \operatorname{F}^{-1}_{\mathcal{N}}(u^l) + \mu^j,     \label{eq:rescaling}
\end{equation}
where $\operatorname{F}^{-1}_{\mathcal{N}}$ is the inverse cumulative distribution function of a standard Gaussian random variable
\begin{equation}
  \operatorname{F}^{-1}_{\mathcal{N}}(u^l) = \sqrt{2}(\operatorname{erf}^{-1}(2 u^l - 1)), \label{eq:standardgaussian}
\end{equation}
and $l = 1,\dots,\dim(\mathcal{N})$,  $\operatorname*{erf}^{-1}$ is the inverse error function, and $L$ is the Cholesky decomposition of the inverse covariance matrix ($\Sigma^{-1} = L^T L$). The Cholesky decomposition exists for positive definite matrices. While only semidefinite generally, the covariance matrices found by the Gaussian mixture model are nondegenerate.
The formula uses the Einstein summation notation.

For multimodal distributions, we introduce a hyperparameter $k=1,\ldots,N_{\textrm{modes}}$ identifying each mode, with prior probability $\prob{k} = m_k$, which is eventually marginalized over.

\section{Sampler Configuration}
\label{sec:sampler_config}

In \Cref{tab:samplerconfig}, we list the configuration of the \texttt{dynesty} sampler used throughout all presented results.

\begin{table}[htbp]
  \centering
  \begin{tabular}{ll}
      \toprule
      Option           & Setting            \\ \midrule
      $\Delta \log Z $ & 0.1                \\
      Bounding         & Multi-ellipsoids   \\
      Sampler          & Random walk        \\
      Walks per iteration  & 25                 \\
      Update interval  & $ 2 n_\text{live}$ \\
      Min. eff.        & $5\%$              \\
      Bootstrap        & 25                 \\
      Enlarge          & 1                  \\ \bottomrule
  \end{tabular}
  \caption{Sampler configuration used in this work. 
      The configuration is oriented at the recommended settings given by the \texttt{dynesty} documentation.
      We use \texttt{dynesty} version~$2.0.3$.
      The reader is referred to its documentation for a detailed description of these options and their effect on the sampling.}
  \label{tab:samplerconfig}
\end{table}

\bibliography{bayesian_fu_method}

@article{Allen:2019vcl,
  title         = {Spherical Ansatz for Parameter-Space Metrics},
  author        = {Allen, Bruce},
  year          = {2019},
  month         = dec,
  journal       = {Phys. Rev. D},
  volume        = {100},
  number        = {12},
  pages         = {124004},
  issn          = {2470-0010, 2470-0029},
  doi           = {10.1103/PhysRevD.100.124004},
  urldate       = {2025-08-17},
  archiveprefix = {arXiv},
  langid        = {english}
}

@article{Arvanitaki:2014wva,
  title         = {Discovering the {{QCD}} Axion with Black Holes and Gravitational Waves},
  author        = {Arvanitaki, Asimina and Baryakhtar, Masha and Huang, Xinlu},
  year          = {2015},
  month         = apr,
  journal       = {Phys. Rev. D},
  volume        = {91},
  number        = {8},
  pages         = {084011},
  issn          = {1550-7998, 1550-2368},
  doi           = {10.1103/PhysRevD.91.084011},
  urldate       = {2024-09-08},
  archiveprefix = {arXiv},
  copyright     = {http://link.aps.org/licenses/aps-default-license},
  langid        = {english}
}

@article{Ashok:2024fts,
  title         = {Bayesian {{F}} -Statistic-Based Parameter Estimation of Continuous Gravitational Waves from Known Pulsars},
  author        = {Ashok, A. and Covas, P. B. and Prix, R. and Papa, M. A.},
  year          = {2024},
  month         = may,
  journal       = {Phys. Rev. D},
  volume        = {109},
  number        = {10},
  pages         = {104002},
  issn          = {2470-0010, 2470-0029},
  doi           = {10.1103/PhysRevD.109.104002},
  urldate       = {2024-09-08},
  archiveprefix = {arXiv},
  langid        = {english}
}

@article{Ashton:2018jfp,
  title      = {Bilby: {{A User-friendly Bayesian Inference Library}} for {{Gravitational-wave Astronomy}}},
  shorttitle = {Bilby},
  year       = {2019},
  month      = apr,
  journal    = {Astrophys. J. Suppl. Ser.},
  volume     = {241},
  number     = {2},
  pages      = {27},
  issn       = {1538-4365},
  doi        = {10.3847/1538-4365/ab06fc},
  urldate    = {2024-06-24},
  author     = {Ashton, Gregory and others}
}

@article{Ashton:2018ure,
  title         = {Hierarchical Multistage {{MCMC}} Follow-up of Continuous Gravitational Wave Candidates},
  author        = {Ashton, G. and Prix, R.},
  year          = {2018},
  month         = may,
  journal       = {Phys. Rev. D},
  volume        = {97},
  number        = {10},
  pages         = {103020},
  issn          = {2470-0010, 2470-0029},
  doi           = {10.1103/PhysRevD.97.103020},
  urldate       = {2024-06-24},
  archiveprefix = {arXiv},
  langid        = {english}
}

@article{Ashton:2022grj,
  title     = {Nested Sampling for Physical Scientists},
  year      = {2022},
  month     = may,
  journal   = {Nat. Revs. Methods Primers},
  volume    = {2},
  number    = {1},
  pages     = {1--22},
  publisher = {Nature Publishing Group},
  issn      = {2662-8449},
  doi       = {10.1038/s43586-022-00121-x},
  urldate   = {2023-10-18},
  copyright = {2022 Springer Nature Limited},
  langid    = {english},
  author    = {Ashton, Greg and others}
}

@article{Beheshtipour:2020nko,
  title         = {Deep Learning for Clustering of Continuous Gravitational Wave Candidates. {{II}}. {{Identification}} of Low-{{SNR}} Candidates},
  author        = {Beheshtipour, B. and Papa, M. A.},
  year          = {2021},
  month         = mar,
  journal       = {Phys. Rev. D},
  volume        = {103},
  number        = {6},
  pages         = {064027},
  issn          = {2470-0010, 2470-0029},
  doi           = {10.1103/PhysRevD.103.064027},
  urldate       = {2024-09-08},
  archiveprefix = {arXiv},
  langid        = {english}
}

@article{Beheshtipour:2020zhb,
  title         = {Deep Learning for Clustering of Continuous Gravitational Wave Candidates},
  author        = {Beheshtipour, B. and Papa, M. A.},
  year          = {2020},
  month         = mar,
  journal       = {Phys. Rev. D},
  volume        = {101},
  number        = {6},
  pages         = {064009},
  issn          = {2470-0010, 2470-0029},
  doi           = {10.1103/PhysRevD.101.064009},
  urldate       = {2024-09-08},
  archiveprefix = {arXiv},
  langid        = {english}
}

@article{Behnke:2014tma,
  author        = {Behnke, Berit and Papa, Maria Alessandra and Prix, Reinhard},
  title         = {{Postprocessing methods used in the search for continuous gravitational-wave signals from the Galactic Center}},
  reportnumber  = {LIGO-DOCUMENT-NUMBER-P1300125, AEI-2014-017},
  doi           = {10.1103/PhysRevD.91.064007},
  journal       = {Phys. Rev. D},
  volume        = {91},
  number        = {6},
  pages         = {064007},
  year          = {2015}
}

@article{Brady:1997ji,
  title         = {Searching for Periodic Sources with {{LIGO}}},
  author        = {Brady, Patrick R. and Creighton, Teviet and Cutler, Curt and Schutz, Bernard F.},
  year          = {1998},
  journal       = {Phys. Rev. D},
  volume        = {57},
  number        = {4},
  pages         = {2101--2116},
  issn          = {0556-2821, 1089-4918},
  doi           = {10.1103/PhysRevD.57.2101},
  urldate       = {2024-09-08},
  archiveprefix = {arXiv},
  copyright     = {http://link.aps.org/licenses/aps-default-license},
  langid        = {english}
}

@article{Brady:1998nj,
  title     = {Searching for Periodic Sources with {{LIGO}}. {{II}}. {{Hierarchical}} Searches},
  author    = {Brady, Patrick R. and Creighton, Teviet},
  year      = {2000},
  month     = feb,
  journal   = {Phys. Rev. D},
  volume    = {61},
  number    = {8},
  pages     = {082001},
  publisher = {American Physical Society},
  doi       = {10.1103/PhysRevD.61.082001},
  urldate   = {2023-07-04}
}

@article{Covas:2024pam,
  title   = {New Framework to Follow up Candidates from Continuous Gravitational-Wave Searches},
  author  = {Covas, P. B. and Prix, R. and Martins, J.},
  year    = {2024},
  month   = jul,
  journal = {Phys. Rev. D},
  volume  = {110},
  number  = {2},
  pages   = {024053},
  issn    = {2470-0010, 2470-0029},
  doi     = {10.1103/PhysRevD.110.024053},
  urldate = {2024-07-29},
  langid  = {english}
}

@article{Cutler:2005hc,
  title      = {Generalized {{F}} -Statistic: {{Multiple}} Detectors and Multiple Gravitational Wave Pulsars},
  shorttitle = {Generalized {{F}} -Statistic},
  author     = {Cutler, Curt and Schutz, Bernard F.},
  year       = {2005},
  month      = sep,
  journal    = {Phys. Rev. D},
  volume     = {72},
  number     = {6},
  pages      = {063006},
  issn       = {1550-7998, 1550-2368},
  doi        = {10.1103/PhysRevD.72.063006},
  urldate    = {2024-06-24},
  copyright  = {http://link.aps.org/licenses/aps-default-license},
  langid     = {english}
}

@article{Dreissigacker:2018afk,
  title   = {Fast and Accurate Sensitivity Estimation for Continuous-Gravitational-Wave Searches},
  author  = {Dreissigacker, Christoph and Prix, Reinhard and Wette, Karl},
  year    = {2018},
  month   = oct,
  journal = {Phys. Rev. D},
  volume  = {98},
  number  = {8},
  pages   = {084058},
  issn    = {2470-0010, 2470-0029},
  doi     = {10.1103/PhysRevD.98.084058},
  urldate = {2024-06-24},
  langid  = {english}
}

@inproceedings{ester:2004,
  title     = {A Density-Based Algorithm for Discovering Clusters in Large Spatial Databases with Noise},
  booktitle = {Proceedings of the Second International Conference on Knowledge Discovery and Data Mining},
  author    = {Ester, Martin and Kriegel, Hans-Peter and Sander, J{\"o}rg and Xu, Xiaowei},
  year      = {1996},
  series    = {{{KDD}}'96},
  pages     = {226--231},
  publisher = {AAAI Press},
  address   = {Portland, Oregon}
}

@misc{GaussianMixtureModels,
  title        = {Gaussian Mixture Models},
  journal      = {scikit-learn},
  note         = {Accessed: 2025-12-08},
  howpublished = {https://scikit-learn.org/stable/modules/mixture.html},
  langid       = {english}
}

@book{gelmanBayesianDataAnalysis2014,
  title     = {Bayesian Data Analysis},
  year      = {2014},
  series    = {Texts in Statistical Science Series},
  edition   = {3},
  publisher = {{CRC Press, Taylor and Francis Group}},
  address   = {Boca Raton London New York},
  isbn      = {978-1-4398-4095-5},
  langid    = {english},
  author    = {Gelman, Andrew and others}
}

@article{Steltner:2021qjy,
  author        = {Steltner, Benjamin and Papa, Maria Alessandra and Eggenstein, Heinz-Bernd},
  title         = {{Identification and removal of non-Gaussian noise transients for gravitational-wave searches}},
  doi           = {10.1103/PhysRevD.105.022005},
  journal       = {Phys. Rev. D},
  volume        = {105},
  number        = {2},
  pages         = {022005},
  year          = {2022}
}

@article{Keitel:2013wga,
  author        = {Keitel, David and Prix, Reinhard and Papa, Maria Alessandra and Leaci, Paola and Siddiqi, Maham},
  title         = {{Search for continuous gravitational waves: Improving robustness versus instrumental artifacts}},
  doi           = {10.1103/PhysRevD.89.064023},
  journal       = {Phys. Rev. D},
  volume        = {89},
  number        = {6},
  pages         = {064023},
  year          = {2014}
}

@article{Jaranowski:1998qm,
  title      = {Data Analysis of Gravitational-Wave Signals from Spinning Neutron Stars: {{The}} Signal and Its Detection},
  shorttitle = {Data Analysis of Gravitational-Wave Signals from Spinning Neutron Stars},
  author     = {Jaranowski, Piotr and Kr{\'o}lak, Andrzej and Schutz, Bernard F.},
  year       = {1998},
  month      = aug,
  journal    = {Phys. Rev. D},
  volume     = {58},
  number     = {6},
  pages      = {063001},
  issn       = {0556-2821, 1089-4918},
  doi        = {10.1103/PhysRevD.58.063001},
  urldate    = {2024-06-24},
  copyright  = {http://link.aps.org/licenses/aps-default-license},
  langid     = {english}
}

@article{Dergachev:2020upb,
  author        = {Dergachev, Vladimir and Papa, Maria Alessandra},
  title         = {{Results from high-frequency all-sky search for continuous gravitational waves from small-ellipticity sources}},
  doi           = {10.1103/PhysRevD.103.063019},
  journal       = {Phys. Rev. D},
  volume        = {103},
  number        = {6},
  pages         = {063019},
  year          = {2021}
}

@article{Dergachev:2011pd,
  author        = {Dergachev, Vladimir},
  title         = {{Loosely coherent searches for sets of well-modeled signals}},
  doi           = {10.1103/PhysRevD.85.062003},
  journal       = {Phys. Rev. D},
  volume        = {85},
  pages         = {062003},
  year          = {2012}
}

@article{Hastings:1970,
  author   = {Hastings, W. K.},
  title    = {Monte Carlo sampling methods using Markov chains and their applications},
  journal  = {Biometrika},
  volume   = {57},
  number   = {1},
  pages    = {97-109},
  year     = {1970},
  month    = {04},
  issn     = {0006-3444},
  doi      = {10.1093/biomet/57.1.97},
}

@article{KAGRA:2021vkt,
  title         = {{{GWTC-3}}: {{Compact Binary Coalescences Observed}} by {{LIGO}} and {{Virgo}} during the {{Second Part}} of the {{Third Observing Run}}},
  shorttitle    = {{{GWTC-3}}},
  year          = {2023},
  month         = oct,
  journal       = {Phys. Rev. X},
  volume        = {13},
  number        = {4},
  pages         = {041039},
  issn          = {2160-3308},
  doi           = {10.1103/PhysRevX.13.041039},
  urldate       = {2024-07-29},
  archiveprefix = {arXiv},
  collaboration = {LIGO Scientific, VIRGO, KAGRA,},
  langid        = {english},
  author        = {Abbott, R. and others}
}

@article{KAGRA:2022dwb,
  title         = {All-Sky Search for Continuous Gravitational Waves from Isolated Neutron Stars Using {{Advanced LIGO}} and {{Advanced Virgo O3}} Data},
  year          = {2022},
  month         = nov,
  journal       = {Phys. Rev. D},
  volume        = {106},
  number        = {10},
  pages         = {102008},
  issn          = {2470-0010, 2470-0029},
  doi           = {10.1103/PhysRevD.106.102008},
  urldate       = {2024-06-26},
  archiveprefix = {arXiv},
  collaboration = {LIGO Scientific, VIRGO, KAGRA,},
  langid        = {english},
  author        = {Abbott, R. and others}
}

@article{KAGRA:2023pio,
  title         = {Open {{Data}} from the {{Third Observing Run}} of {{LIGO}}, {{Virgo}}, {{KAGRA}}, and {{GEO}}},
  year          = {2023},
  month         = jul,
  journal       = {Astrophys. J. Suppl. Ser.},
  volume        = {267},
  number        = {2},
  pages         = {29},
  publisher     = {The American Astronomical Society},
  issn          = {0067-0049},
  doi           = {10.3847/1538-4365/acdc9f},
  urldate       = {2023-10-23},
  archiveprefix = {arXiv},
  collaboration = {LIGO Scientific, VIRGO, KAGRA,},
  langid        = {english},
  author        = {Abbott, R. and others}
}

@misc{known-lines-list,
  author	="Goetz, Evan and others",
  year	= "2021",
  title        = {{{LIGO-T2100200-v2}}: {{O3}} Lines and Combs in Found in Self-Gated {{C01}} Data},
  note         = {Date accessed: 2025-08-06},
  howpublished = {\url{https://dcc.ligo.org/LIGO-T2100200/public}}
}

@article{Krishnan:2004sv,
  title         = {Hough Transform Search for Continuous Gravitational Waves},
  year          = {2004},
  journal       = {Phys. Rev. D},
  volume        = {70},
  number        = {8},
  pages         = {082001},
  issn          = {1550-7998, 1550-2368},
  doi           = {10.1103/PhysRevD.70.082001},
  urldate       = {2024-09-08},
  archiveprefix = {arXiv},
  copyright     = {http://link.aps.org/licenses/aps-default-license},
  langid        = {english},
  author        = {Krishnan, Badri and others}
}

@misc{lalsuite,
       author         = "{LIGO Scientific Collaboration} and {Virgo Collaboration} and {KAGRA Collaboration}",
       title          = "{LVK} {A}lgorithm {L}ibrary - {LALS}uite",
       howpublished   = "Free software (GPL)",
       doi            = "10.7935/GT1W-FZ16",
       year           = "2018"
 }

@article{LIGOScientific:2020qhb,
  title         = {All-Sky Search in Early {{O3 LIGO}} Data for Continuous Gravitational-Wave Signals from Unknown Neutron Stars in Binary Systems},
  year          = {2021},
  month         = mar,
  journal       = {Phys. Rev. D},
  volume        = {103},
  number        = {6},
  pages         = {064017},
  issn          = {2470-0010, 2470-0029},
  doi           = {10.1103/PhysRevD.103.064017},
  urldate       = {2024-07-23},
  archiveprefix = {arXiv},
  collaboration = {LIGO Scientific, VIRGO},
  langid        = {english},
  author        = {Abbott, R. and others}
}

@article{LIGOScientific:2021hvc,
  title         = {Searches for {{Gravitational Waves}} from {{Known Pulsars}} at {{Two Harmonics}} in the {{Second}} and {{Third LIGO-Virgo Observing Runs}}},
  year          = {2022},
  month         = may,
  journal       = {Astrophys. J.},
  volume        = {935},
  number        = {1},
  pages         = {1},
  issn          = {0004-637X, 1538-4357},
  doi           = {10.3847/1538-4357/ac6acf},
  urldate       = {2024-09-08},
  archiveprefix = {arXiv},
  collaboration = {LIGO Scientific, VIRGO, KAGRA},
  author        = {Abbott, R. and others}
}

@article{LIGOScientific:2024elc,
  title         = {Observation of {{Gravitational Waves}} from the {{Coalescence}} of a 2.5--4.5 {{M}} {\textsubscript{{$\odot$}}} {{Compact Object}} and a {{Neutron Star}}},
  year          = {2024},
  month         = jul,
  journal       = {Astrophys. J. Lett.},
  volume        = {970},
  number        = {2},
  pages         = {L34},
  issn          = {2041-8205, 2041-8213},
  doi           = {10.3847/2041-8213/ad5beb},
  urldate       = {2024-08-27},
  archiveprefix = {arXiv},
  collaboration = {LIGO Scientific, VIRGO, KAGRA},
  author        = {Abac, A. G. and others}
}

@misc{ligo-o4-observing-plan-current,
  author      = {Shoemaker, David and Arnaud, Nicolas and Sawada, Takahiro},
  year         = {2025},
  month        = {nov},
  day          = {18},
  title        = {{{LIGO-G2002127-v33}}: {{Observing Scenario}} Timeline Graphic, Post-{{O3}}},
  note         = {Date accessed: 2025-12-08},
  howpublished = {https://dcc.ligo.org/LIGO-G2002127/public}
}

@article{Mirasola:2024lcq,
  title         = {Toward a Computationally Efficient Follow-up Pipeline for Blind Continuous Gravitational-Wave Searches},
  author        = {Mirasola, Lorenzo and Tenorio, Rodrigo},
  year          = {2024},
  month         = dec,
  journal       = {Phys. Rev. D},
  volume        = {110},
  number        = {12},
  pages         = {124049},
  issn          = {2470-0010, 2470-0029},
  doi           = {10.1103/PhysRevD.110.124049},
  urldate       = {2025-08-14},
  archiveprefix = {arXiv},
  langid        = {english}
}

@article{Ming:2024duga,
  title = {Deep {{Einstein}}@{{Home Search}} for {{Continuous Gravitational Waves}} from the {{Central Compact Objects}} in the {{Supernova Remnants Vela Jr}}. and {{G347}}.3-0.5 {{Using LIGO Public Data}}},
  year = {2024},
  month = dec,
  journal = {Astrophys. J.},
  volume = {977},
  number = {2},
  pages = {154},
  publisher = {The American Astronomical Society},
  issn = {0004-637X},
  doi = {10.3847/1538-4357/ad8b9e},
  urldate = {2025-08-29},
  langid = {english},
  author = {Ming, J. and others}
}

@article{Messenger:2008ta,
  title         = {Random Template Banks and Relaxed Lattice Coverings},
  author        = {Messenger, C. and Prix, R. and Papa, M. A.},
  year          = {2009},
  journal       = {Phys. Rev. D},
  volume        = {79},
  number        = {10},
  pages         = {104017},
  issn          = {1550-7998, 1550-2368},
  doi           = {10.1103/PhysRevD.79.104017},
  urldate       = {2025-08-13},
  archiveprefix = {arXiv},
  copyright     = {http://link.aps.org/licenses/aps-default-license},
  langid        = {english}
}

@book{numericalrecipes,
  author    = {Press, William H. and Teukolsky, Saul A. and Vetterling, William T. and Flannery, Brian P.},
  title     = {Numerical Recipes 3rd Edition: The Art of Scientific Computing},
  year      = {2007},
  isbn      = {0521880688},
  publisher = {Cambridge University Press},
  address   = {USA},
  edition   = {3}
}

@article{Pagliaro:2023bvi,
  title      = {Continuous {{Gravitational Waves}} from {{Galactic Neutron Stars}}: {{Demography}}, {{Detectability}}, and {{Prospects}}},
  shorttitle = {Continuous {{Gravitational Waves}} from {{Galactic Neutron Stars}}},
  year       = {2023},
  month      = aug,
  journal    = {Astrophys. J.},
  volume     = {952},
  number     = {2},
  pages      = {123},
  issn       = {0004-637X, 1538-4357},
  doi        = {10.3847/1538-4357/acd76f},
  urldate    = {2024-06-24},
  author     = {Pagliaro, Gianluca and others}
}

@article{Papa:2016cwb,
  title         = {Hierarchical Follow-up of Subthreshold Candidates of an All-Sky {{Einstein}}@{{Home}} Search for Continuous Gravitational Waves on {{LIGO}} Sixth Science Run Data},
  year          = {2016},
  month         = dec,
  journal       = {Phys. Rev. D},
  volume        = {94},
  number        = {12},
  pages         = {122006},
  issn          = {2470-0010, 2470-0029},
  doi           = {10.1103/PhysRevD.94.122006},
  urldate       = {2024-09-08},
  archiveprefix = {arXiv},
  copyright     = {http://creativecommons.org/licenses/by/3.0/},
  langid        = {english},
  author        = {Papa, Maria Alessandra and others}
}

@article{Pedregosa:2011ork,
  author        = {Pedregosa, Fabian and others},
  title         = {{Scikit-learn: Machine Learning in Python}},
  eprint        = {1201.0490},
  archiveprefix = {arXiv},
  primaryclass  = {cs.LG},
  journal       = {J. Machine Learning Res.},
  volume        = {12},
  number = {85},
  pages         = {2825--2830},
  year          = {2011}
}

@article{Prix:2006wm,
  title      = {Search for Continuous Gravitational Waves: {{Metric}} of the Multidetector {{F}} -Statistic},
  shorttitle = {Search for Continuous Gravitational Waves},
  author     = {Prix, Reinhard},
  year       = {2007},
  month      = jan,
  journal    = {Phys. Rev. D},
  volume     = {75},
  number     = {2},
  pages      = {023004},
  issn       = {1550-7998, 1550-2368},
  doi        = {10.1103/PhysRevD.75.023004},
  urldate    = {2023-06-13},
  langid     = {english}
}

@article{Riles:2022wwz,
  title   = {Searches for Continuous-Wave Gravitational Radiation},
  author  = {Riles, Keith},
  year    = {2023},
  month   = apr,
  journal = {Living Reviews in Relativity},
  volume  = {26},
  number  = {1},
  pages   = {3},
  issn    = {1433-8351},
  doi     = {10.1007/s41114-023-00044-3},
  urldate = {2023-11-16},
  langid  = {english}
}

@misc{Schutz:1999mb,
  title         = {End-to-End Algorithm for Hierarchical Area Searches for Long-Duration {{GW}} Sources for {{GEO}} 600},
  author        = {Schutz, Bernard F. and Papa, M. Alessandra},
  year          = {2000},
  urldate       = {2024-09-08},
  eprint        = {gr-qc/9905018},
  archiveprefix = {arXiv}
}

@misc{Searle:2008jv,
  title     = {Monte-{{Carlo}} and {{Bayesian}} Techniques in Gravitational Wave Burst Data Analysis},
  author    = {Searle, Antony C.},
  year      = {2008},
  archiveprefix = {arXiv},
  eprint = {gr-qc/0804.1161},
  urldate   = {2024-07-26},
}

@article{Skilling:2004pqw,
  title   = {Nested {{Sampling}}},
  author  = {Skilling, John},
  year    = {2004},
  month   = nov,
  journal = {AIP Conference Proceedings},
  volume  = {735},
  number  = {1},
  pages   = {395--405},
  issn    = {0094-243X},
  doi     = {10.1063/1.1835238},
  urldate = {2023-10-23}
}

@article{Speagle:2019ivv,
  title      = {Dynesty: A Dynamic Nested Sampling Package for Estimating {{Bayesian}} Posteriors and Evidences},
  shorttitle = {Dynesty},
  author     = {Speagle, Joshua S},
  year       = {2020},
  month      = apr,
  journal    = {Mon. Not. R. Astron. Soc.},
  volume     = {493},
  number     = {3},
  pages      = {3132--3158},
  issn       = {0035-8711, 1365-2966},
  doi        = {10.1093/mnras/staa278},
  urldate    = {2024-06-24},
  copyright  = {https://academic.oup.com/journals/pages/open\_access/funder\_policies/chorus/standard\_publication\_model},
  langid     = {english}
}

@article{Steltner:2022aze,
  title         = {Density-Clustering of Continuous Gravitational Wave Candidates from Large Surveys},
  author        = {Steltner, B. and Menne, T. and Papa, M. A. and Eggenstein, H.-B.},
  year          = {2022},
  month         = nov,
  journal       = {Phys. Rev. D},
  volume        = {106},
  number        = {10},
  pages         = {104063},
  issn          = {2470-0010, 2470-0029},
  doi           = {10.1103/PhysRevD.106.104063},
  urldate       = {2024-09-08},
  archiveprefix = {arXiv},
  langid        = {english}
}

@article{Steltner:2023cfk,
  title   = {Deep {{Einstein}}@{{Home All-sky Search}} for {{Continuous Gravitational Waves}} in {{LIGO O3 Public Data}}},
  year    = {2023},
  month   = jul,
  journal = {Astrophys. J.},
  volume  = {952},
  number  = {1},
  pages   = {55},
  issn    = {0004-637X, 1538-4357},
  doi     = {10.3847/1538-4357/acdad4},
  urldate = {2024-07-29},
  author  = {Steltner, B. and others}
}

@article{Tenorio:2021njf,
  title   = {Application of a Hierarchical {{MCMC}} Follow-up to {{Advanced LIGO}} Continuous Gravitational-Wave Candidates},
  author  = {Tenorio, Rodrigo and Keitel, David and Sintes, Alicia M.},
  year    = {2021},
  month   = oct,
  journal = {Phys. Rev. D},
  volume  = {104},
  number  = {8},
  pages   = {084012},
  issn    = {2470-0010, 2470-0029},
  doi     = {10.1103/PhysRevD.104.084012},
  urldate = {2023-10-18},
  langid  = {english}
}

@article{Wette:2015lfa,
  title   = {Parameter-Space Metric for All-Sky Semicoherent Searches for Gravitational-Wave Pulsars},
  author  = {Wette, Karl},
  year    = {2015},
  month   = oct,
  journal = {Phys. Rev. D},
  volume  = {92},
  number  = {8},
  pages   = {082003},
  issn    = {1550-7998, 1550-2368},
  doi     = {10.1103/PhysRevD.92.082003},
  urldate = {2024-01-19},
  langid  = {english}
}

@article{Wette:2020air,
  title      = {{{SWIGLAL}}: {{Python}} and {{Octave}} Interfaces to the {{LALSuite}} Gravitational-Wave Data Analysis Libraries},
  shorttitle = {{{SWIGLAL}}},
  author     = {Wette, Karl},
  year       = {2020},
  month      = jul,
  journal    = {SoftwareX},
  volume     = {12},
  pages      = {100634},
  issn       = {23527110},
  doi        = {10.1016/j.softx.2020.100634},
  urldate    = {2024-06-24},
  langid     = {english}
}

@article{Zhu:2020tht,
  title         = {Characterizing the Continuous Gravitational-Wave Signal from Boson Clouds around {{Galactic}} Isolated Black Holes},
  year          = {2020},
  month         = sep,
  journal       = {Phys. Rev. D},
  volume        = {102},
  number        = {6},
  pages         = {063020},
  issn          = {2470-0010, 2470-0029},
  doi           = {10.1103/PhysRevD.102.063020},
  urldate       = {2024-09-08},
  archiveprefix = {arXiv},
  langid        = {english},
  author        = {Zhu, Sylvia J. and others}
}

@article{prixSearchMethodLongduration2011a,
  title = {Search Method for Long-Duration Gravitational-Wave Transients from Neutron Stars},
  author = {Prix, R. and Giampanis, S. and Messenger, C.},
  year = {2011},
  month = jul,
  journal = {Phys. Rev. D},
  volume = {84},
  number = {2},
  pages = {023007},
  issn = {1550-7998, 1550-2368},
  doi = {10.1103/PhysRevD.84.023007},
  urldate = {2023-01-13},
  langid = {english}
}

@article{covas2025,
  title = {Generalized Parameter-Space Metrics for Continuous Gravitational-Wave Searches},
  author = {Covas, P. B. and Prix, R.},
  year = {2025},
  month = aug,
  journal = {Phys. Rev. D},
  volume = {112},
  number = {4},
  pages = {044041},
  publisher = {American Physical Society},
  doi = {10.1103/1vds-2cgh},
  urldate = {2025-08-22}
}

@article{Covas:2018oik,
	author = "Covas, P.B. and others",
	collaboration = "LSC",
	title = "{Identification and mitigation of narrow spectral artifacts that degrade searches for persistent gravitational waves in the first two observing runs of Advanced LIGO}",
	reportNumber = "LIGO-P1700440",
	doi = "10.1103/PhysRevD.97.082002",
	journal = "Phys. Rev. D",
	volume = "97",
	number = "8",
	pages = "082002",
	year = "2018"
}

@article{Davis:2018yrz,
	author = "Davis, D. and Massinger, T. J. and Lundgren, A. P. and Driggers, J. C. and Urban, A. L. and Nuttall, L. K.",
	title = "{Improving the Sensitivity of Advanced LIGO Using Noise Subtraction}",
	doi = "10.1088/1361-6382/ab01c5",
	journal = "Classical and Quantum Gravity",
	volume = "36",
	number = "5",
	pages = "055011",
	year = "2019"
}

@article{LIGO:2021ppb,
	author = "Davis, Derek and others",
	collaboration = "LIGO",
	title = "{LIGO detector characterization in the second and third observing runs}",
	reportNumber = "P2000495",
	doi = "10.1088/1361-6382/abfd85",
	journal = "Classical and Quantum Gravity",
	volume = "38",
	number = "13",
	pages = "135014",
	year = "2021"
}

@article{Capote:2024rmo,
	author = "Capote, E. and others",
	title = "{Advanced LIGO detector performance in the fourth observing run}",
	primaryClass = "gr-qc",
	reportNumber = "LIGO-P2400256",
	doi = "10.1103/PhysRevD.111.062002",
	journal = "Phys. Rev. D",
	volume = "111",
	number = "6",
	pages = "062002",
	year = "2025"
}

@article{Covas:2024nzs,
  title = {Search for {{Continuous Gravitational Waves}} from {{Unknown Neutron Stars}} in {{Binary Systems}} with {{Long Orbital Periods}} in {{O3 Data}}},
  author = {Covas, P. B. and Papa, M. A. and Prix, R.},
  year = {2025},
  month = may,
  journal = {Astrophys. J.},
  volume = {985},
  number = {2},
  pages = {192},
  issn = {0004-637X, 1538-4357},
  doi = {10.3847/1538-4357/adcc2b},
  urldate = {2025-08-22},
  archiveprefix = {arXiv},
  langid = {english}
}

@misc{Dergachev:2025ead,
    author = "Dergachev, Vladimir and Papa, Maria Alessandra",
    title = "{Early release of low-frequency atlas of continuous gravitational waves}",
    eprint = "2507.12161",
    archivePrefix = "arXiv",
    primaryClass = "gr-qc",
    month = "7",
    year = "2025"
}

@misc{McGloughlin:2025eso,
    author = "McGloughlin, Brian and Steltner, Benjamin and Martins, Jasper and Papa, Maria Alessandra and Eggenstein, Heinz-Bernd and Ming, Jing and Machenschalk, Bernd and Prix, Reinhard and Bensch, Maximillian",
    title = "{High-frequency continuous gravitational waves searched in LIGO O3 public data with Einstein@Home}",
    eprint = "2508.20073",
    archivePrefix = "arXiv",
    primaryClass = "gr-qc",
    month = "8",
    year = "2025"
}

@article{McGloughlin:2025iyx,
  title = {Einstein@{{Home All-sky}} ``{{Bucket}}'' {{Search}} for {{Continuous Gravitational Waves}} in {{LIGO O3a Public Data}}},
  year = 2026,
  month = jan,
  journal = {Astrophys. J.},
  volume = {997},
  number = {2},
  pages = {149},
  issn = {0004-637X, 1538-4357},
  doi = {10.3847/1538-4357/ae225a},
  urldate = {2026-01-26},
  archiveprefix = {arXiv},
  author = {McGloughlin, B. and others}
}

\end{document}